\documentclass[12pt]{article}
\pdfoutput=1
\usepackage{amssymb,cite,graphicx}
\usepackage{amsmath}
\usepackage{amsfonts}
\usepackage[titletoc,title]{appendix}
\usepackage{bbold}
\usepackage[small]{caption}
\usepackage[margin=1in]{geometry}
\usepackage[multiple]{footmisc}
\usepackage{mathtools}
\usepackage{physics}
\usepackage{slashed}
\usepackage{xcolor}

\definecolor{mediumblue}{rgb}{0,0,0.8}
\usepackage{hyperref}
\hypersetup{%
  linktocpage=true,
  colorlinks=true,
  citecolor=mediumblue,
  filecolor=mediumblue,
  linkcolor=mediumblue,
  urlcolor=mediumblue,
}

\graphicspath{{./}{./figures/}}
\numberwithin{equation}{section}
\allowdisplaybreaks%

\newcommand{\diag}{\mathop{\mathrm{diag}}\nolimits}

\def\thefootnote{\fnsymbol{footnote}}

\newcommand{\be}{\begin{equation}}
\newcommand{\ee}{\end{equation}}
\newcommand{\bea}{\begin{eqnarray}}
\newcommand{\eea}{\end{eqnarray}}

\def\thefootnote{\fnsymbol{footnote}}

\begin{document}

\begin{titlepage}
  \begin{flushright}
   \texttt{CTPU-PTC-20-19}
  \end{flushright}
  \medskip

  \begin{center}
    {\Large\bf\boldmath
      Flavor and CP-violating Higgs sector \vspace{0.2cm}\\
      in two Higgs doublet models with $U(1)'$}
\vspace{1.5cm}

    {\bf Ligong~Bian$^{1,*}$, Hyun~Min~Lee$^{2,\dagger}$ and Chan~Beom~Park$^{3,\ddagger}$}

    \medskip

    {\it $^1$Department of Physics,
      Chongqing University, Chongqing 401331, China}\\[0.2cm]
    {\it $^2$Department of Physics, Chung-Ang University, Seoul
      06974, Korea}\\[0.2cm]
    {\it $^3$Center for Theoretical Physics of the Universe,\\
      Institute for Basic Science (IBS), Daejeon 34126, Korea}
  \end{center}

  \bigskip

  \begin{abstract}
    \noindent
    We investigate the role of a local $U(1)'$ symmetry for the problem of CP violation in the effective theory for two Higgs doublet models and its microscopic counterparts.  First, in two Higgs doublet models with $U(1)'$, we show that the higher-dimensional operators in the scalar potential violate the CP symmetry with an interplay with the mixing mass parameter, giving rise to small mixings between CP-even and CP-odd scalars. Motivated by the $B$-meson anomalies in recent years, we take the flavored $U(1)'$ to be a benchmark model for specifying the flavor structure. Then, we calculate the electric dipole moment of electron (eEDM) at two loops due to the CP-violating higher-dimensional operators and identify the correlation between the masses of heavy Higgs bosons and the cutoff scale from the bound on eEDM. We also comment on the possibility of making an independent test of the CP violation in the collider searches for heavy Higgs bosons. Finally, we show how the obtained eEDM results in the effective theory can be used to constrain microscopic models with an explicit CP violation in the partially decoupled or dark sectors.
  \end{abstract}

  \vspace*{\fill}
  \begin{flushleft}
    $^*$Email: lgbycl@cqu.edu.cn  \\
    $^\dagger$Email: hminlee@cau.ac.kr \\
    $^\ddagger$Email: cbpark@ibs.re.kr
  \end{flushleft}
\end{titlepage}

\renewcommand{\thefootnote}{\arabic{footnote}}
\setcounter{footnote}{0}

\section{Introduction}

It is an important task to understand the flavor structure of the Standard Model (SM) and the origin of CP violation, calling for physics beyond the SM\@.
In particular, a violation of lepton flavor universality would be an important indirect test of the SM, and it provides a guideline for going beyond the SM and designing the high energy colliders in the next generations beyond the Large Hadron Collider (LHC).
We also need to look for the source of a new CP violation in order to explain the matter-antimatter asymmetry in the Universe.

In models with an extended Higgs sector, we may have new sources for
the CP violation unless CP is conserved by a symmetry argument or ansatz~\cite{2HDM}.
One of the most stringent constraints on the CP violation is from the
electric dipole moment (EDM) of the electron and the neutron
counterparts.
Thus, we need to find a way to make a sufficient suppression of the new physics contributions of new CP phases to the EDMs.
In this regard, an extra $U(1)'$ symmetry plays an important role in controlling the CP violation, at least, in the Higgs sector, because it protects the CP symmetry from being broken at the renormalizable level in two Higgs doublet models (2HDMs).

In recent years, there have been interesting anomalies in the semi-leptonic decays of $B$-mesons, hinting at the violation of lepton flavor universality in the SM with about $2$--$3\sigma$ deviation at each observable.
The measured values of $R_K={\cal B}(B\rightarrow K\mu^+\mu^-)/{\cal
  B}(B\rightarrow Ke^+e^-)$ from LHCb data~\cite{RK,RK-new} as well as the similar ratio for vector $B$-mesons, $R_{K^*}={\cal B}(B\rightarrow K^*\mu^+\mu^-)/{\cal B}(B\rightarrow
K^*e^+e^-)$ from LHCb~\cite{RKs}, show deviations from the SM predictions.
The deviation in $R_{K^*}$ is supported by the discrepancy in the angular distribution of $B\rightarrow K^*\mu^+\mu^-$~\cite{P5}. The recent update on $R_{K^*}$ from Belle data~\cite{RKs-new} shows a similar deviation in low energy bins, although the combined fits in high energy bins in Belle~\cite{altman} are consistent with the SM but with large error bars.
As a result, the combined significance for the global fits of the $B$-meson data turns out to be about 5--$6\sigma$~\cite{RK-newfit,update}.

The $B$-meson anomalies can be accounted for by the introduction of a flavor-dependent $U(1)'$ distinguishing between leptons in the SM~\cite{Bian:2017rpg,Bian:2017xzg}. But, the flavor $U(1)'$ necessarily requires at least two Higgs doublets, and it gives rise to flavor-violating couplings for the $Z'$ gauge boson and new Higgs bosons~\cite{Bian:2017xzg}.
As a consequence, there are testable signatures of the flavor-dependent $U(1)'$ from other $B$-meson decays and mixings, as well as  flavor-violating productions and decays of heavy Higgs bosons at the LHC~\cite{Bian:2017xzg}.
However, in this class of models with flavored $U(1)'$, the CP symmetry is well protected at the renormalizable level.
In order to induce new CP phases without extra particles at low
energy, it is inevitable to go beyond the renormalizable level and
include higher-dimensional operators in the scalar potential.
Then, we can use the experimental result of the EDMs to set a bound on the scale of
new physics responsible for the CP violation.
The required higher-dimensional operators can be originated from several UV-complete
models such as the $U(1)'$-symmetric Next-to-Minimal
Supersymmetric Standard Model (NMSSM), where the scale of new physics for CP violation
is set by the mass parameters of heavy top squarks.

In this article, taking a flavor-dependent $U(1)'$ as a benchmark
model in Refs.~\cite{Bian:2017rpg,Bian:2017xzg} to explain the
$B$-meson anomalies, we undertake a concrete discussion on the problem
of CP violation in the 2HDM\@. We investigate the salient features of
the CP-violating Higgs sector in the effective theory that are
applicable to general 2HDMs with $U(1)'$ beyond any flavor-specific
$U(1)'$,
as far as there is no significant flavor violation in the Yukawa
couplings for charged leptons.
Including higher-dimensional operators with extra CP phases in the Higgs potential, we show that the mixing mass parameter in the Higgs potential gets a nontrivial CP phase by the tadpole condition, resulting in the mixings between CP-even and CP-odd neutral scalars in
the model. Taking into account the contribution of the new CP phase to the EDM of the electron, we correlate between heavy Higgs boson masses and new physics scales for the higher-dimensional operators. We also address the implications of the Higgs mixings for the collider searches of CP-violating Higgs resonances and present some microscopic $U(1)'$ models with extra matter content for generating the higher-dimensional operators with new CP phases in the Higgs potential.

The rest of the paper is organized as follows.
We begin with a description of the Higgs potential in two Higgs doublet models with a local $U(1)'$ and take the $U(1)'$ to be flavor-dependent for the specific Yukawa couplings for the SM fermions. Then, we study the tadpole conditions from the minimization of the potential in the presence of the $U(1)'$ invariant higher-dimensional operators up to dimension-6 and derive the mixings between neutral scalars of the model and the CP-violating Yukawa couplings to SM quarks and leptons.
Next, we update the constraints on the $Z'$ mass and couplings from $B$-meson anomalies and calculate the EDMs from the Higgs mixings. We also comment on the anomalous magnetic dipole moment of leptons and the collider searches for CP-violating resonance searches for extra Higgs bosons.
We continue to provide the NMSSM and the $U(1)'$ models with extra doublet and singlet scalars or fermions as the microscopic origin of the higher-dimensional operators with nontrivial CP phases.
Finally, conclusions are drawn.
There are four appendices, dealing with the minimization conditions, the diagonalization of scalar mass matrices, the diagonalization of quark mass matrices, and the self-interactions and gauge interactions of scalar fields.

\section{Two Higgs doublet models with local \boldmath{$U(1)'$}}

In models with a local $U(1)'$ under which two Higgs doublets carry nonzero charges,\footnote{We can always make an overall shift of $U(1)'$ charges such that one of the Higgs doublets is neutral under $U(1)'$, without loss of generality. This will be the case in a concrete model for flavored $U(1)'$ in the later discussion.} we consider the CP symmetry in the scalar potential and describe the flavored $U(1)'$ for a concrete discussion on the Yukawa structure in this model.

\subsection{\boldmath The scalar potential with $U(1)'$}

The Higgs sector of the SM is CP conserving at the renormalizable
level.
One may attempt to extend the Higgs sector to accommodate CP violation
by including additional matters or introducing higher-dimensional
operators, or both.
One of the simplest ways to induce CP violation is to add
one more Higgs doublet as in the 2HDM~\cite{2HDM}.
The Higgs potential in the most general 2HDM includes the terms of
\begin{equation}
  V_\text{2HDM} \supset
  -\mu^2 H_1^\dagger H_2 + \lambda_5 (H_1^\dagger H_2)^2
  + \lambda_6 (H_1^\dagger H_1) (H_1^\dagger H_2)
  + \lambda_7 (H_2^\dagger H_2) (H_1^\dagger H_2) + \mathrm{h.c.} ,
  \label{eq:2HDM_potential}
\end{equation}
where the $\mu$ parameter and the quartic couplings $\lambda_i$ are
complex.
Even if one imposes a softly broken $Z_2$ symmetry to forbid
tree-level flavor-changing neutral currents, the $\mu$ and $\lambda_5$
terms remain, thus enabling us to accommodate CP violation at tree level.

Let us consider an extra $U(1)'$ and suppose that the two Higgs
doublets are charged differently under the $U(1)'$:
$Q_{H_1}' \neq Q_{H_2}'$.
Then, it is straightforward to see that all the terms that can
induce CP violation in~(\ref{eq:2HDM_potential}) are forbidden.
Therefore, we can argue that CP is an accidental symmetry in the
presence of such $U(1)'$.

For a realistic model, we further introduce a complex singlet scalar
$S$ responsible for breaking the $U(1)'$ symmetry spontaneously and
allowing for a correct electroweak symmetry breaking.
As a result, in terms of the primed notations for scalar fields and
their couplings for convenience in the later discussion, the
renormalizable scalar potential is given by
\begin{align}
 V_1=
  &~\mu^{\prime 2}_1 |H'_1|^2 + \mu^{\prime 2}_2 |H'_2|^2- \left( \mu' S H^{'\dagger}_1 H'_2
    + \mathrm{h.c.}\right) \nonumber \\
  &+\lambda'_1 |H'_1|^4+\lambda'_2 |H'_2|^4 + 2\lambda_3
    |H'_1|^2|H'_2|^2+2\lambda'_4 (H^{\prime\dagger}_1 H'_2)(H^{\prime \dagger}_2 H'_1)
    \nonumber \\
  &+ 2 |S'|^2(\kappa'_1 |H'_1|^2 +\kappa'_2
    |H'_2|^2)+m^{\prime 2}_{S}|S'|^2+\lambda'_{S}|S'|^4.\label{s1}
\end{align}
Here $\mu'$ is the only complex parameter.
It turns out that the extra CP phase coming from $\mu'$ can be set
zero due to the tadpole conditions, as shown in
Appendix~\ref{app:higgs_sector}.
Therefore, CP is still an accidental symmetry at the renormalizable level.
We note that
the consequence generally holds for the 2HDMs where the Higgs doublets carry different $U(1)'$ charges, irrespective of whether the $U(1)'$ symmetry is flavor-dependent or not.

A way out for having CP violation in the Higgs sector is to include
higher-dimensional operators that can generically give rise to nonzero
CP phases.
The operators are required to be nonvanishing even when we impose the
tadpole conditions.
In addition to the renormalizable scalar potential in Eq.~(\ref{s1}),
we can add the higher-dimensional terms respecting the $U(1)'$ gauge
invariance as follows:
\begin{equation}
  V_2=\frac{c'_1}{\Lambda^2}\,
  S^{\prime 2} (H^{\prime\dagger}_1 H'_2)^2 + \mathrm{h.c.}
  + \cdots , \label{s30}
\end{equation}
where $c_1^\prime$ is complex
and the ellipses denote even higher-dimensional terms.
Then, as will be discussed in the next section, there appears a
nontrivial CP phase from the higher-dimensional operator, leading to
the mixings between CP-even and CP-odd scalars in the general 2HDM
with $U(1)'$.\footnote{We note that there can also be
  dimension-5 interactions for the Yukawa couplings which potentially
  carry extra CP phases, but these are model-dependent.}
Then the questions worth investigating are what the size of $\Lambda$
and its UV origin could be.
Currently, we find that one of the best motivations for the extra
$U(1)'$ from the phenomenological perspective is from the possibility
of explaining the $B$-meson anomalies at LHCb.
Thus, we take the $U(1)'$ to couple to the SM fermions in a
flavor-dependent way and determine the Yukawa couplings with
flavor-dependent $U(1)'$.
It enables us to make a concrete discussion of the CP violation using
physical observables.
In the next subsection and the sections that follow, we examine the
Higgs and fermions sectors with a flavored $U(1)'$.
Then, in Sec.~\ref{sec:EDM_collider}, we show the constraints on
$\Lambda$ from the current eEDM result and discuss the collider
signatures that can be probed at the LHC and future colliders.
The viable origins of higher-dimensional operators are presented in
Sec.~\ref{sec:UV}.

\subsection{\boldmath A concrete model for flavored $U(1)'$}

For a concrete discussion on the violation of flavor and CP, we henceforth take a specific model for the flavored $U(1)'$.
Nonetheless, we stress that our following discussion on the CP-violating Higgs still holds for general 2HDMs with $U(1)'$, under which either of two Higgs doublets carries a nonzero charge.

We regard the new gauge boson $Z^\prime$ associated with the $U(1)'$ symmetry to couple specifically to heavy flavors  as a linear combination of ${U(1)}_{L_\mu-L_\tau}$ and
${U(1)}_{B_3-L_3}$ as follows:
\begin{equation}
  Q' \equiv y(L_\mu-L_\tau)+x(B_3-L_3) ,
\end{equation}
where $x$ and $y$ are real parameters.\footnote{
In our setup, $L_\mu - L_\tau$ and $L_2 - L_3$ can be used
interchangeably.
We note that there are a lot of similar models with flavor-dependent
$U(1)'$, including either $U(1)_{B_3-L_3}$~\cite{b3l3} or
$U(1)_{L_\mu-L_\tau}$~\cite{lmultau}, with a motivation to explain the
$B$-meson anomalies at LHCb.
}
Here only the ratio of the $x$ and
$y$ parameters is physically meaningful as either of them is absorbed
by the $Z'$ gauge coupling.
Then, it is necessary to introduce two Higgs doublets $H_{1,2}$ for obtaining correct quark masses and mixings~\cite{Bian:2017rpg,Bian:2017xzg}.
Moreover, in order to cancel the gauge anomalies,
the fermion sector is required to include at least two right-handed neutrinos
$\nu_{iR}$ ($i = 2$, $3$).
One more right-handed neutrino $\nu_{1R}$ with zero $U(1)'$ charge as well as
extra singlet scalars, $\Phi_a$ $(a=1,\,2,\,3)$, with $U(1)'$ charges
of $-y$, $x+y$, $x$, respectively, are also necessary for neutrino
masses and mixings, in addition to the two Higgs doublets and the
complex singlet scalar.
The $U(1)'$ charge assignments are given in Table~\ref{modelA}.
\begin{table}[hbt!]
  \begin{center}
    \begin{tabular}{c|ccccccccc}
      \hline\hline
      &&&&&&&&&\\[-2mm]
      & $q_{3L}$ & $u_{3R}$  &  $d_{3R}$ & $\ell_{2L}$  & $e_{2R}$
      & $\nu_{2R}$ & $\ell_{3L}$ & $e_{3R}$ & $\nu_{3R}$\\[2mm]
      \hline
      &&&&&&&&&\\[-2mm]
      $Q'$ & $\frac{1}{3}x$ & $\frac{1}{3}x$ & $\frac{1}{3}x$
            & $y$ & $y$ & $y$ & $-x-y$ & $-x-y$ & $-x-y$\\[2mm]
      \hline\hline
    \end{tabular}\\[2mm]
    \begin{tabular}{c|cccccc}
      \hline\hline
      &&&&&&\\[-2mm]
      & $S$  & $H_1$ & $H_2$ & $\Phi_1$ & $\Phi_2$ & $\Phi_3$\\[2mm]
      \hline
      &&&&&&\\[-2mm]
      $Q'$ &  $\frac{1}{3}x$ & $0$ & $-\frac{1}{3}x$  & $-y$ & $x+y$
                     & $x$\\[2mm]
      \hline\hline
    \end{tabular}
  \end{center}
    \caption{$U(1)'$ charges of fermions and scalars.\label{modelA}}
\end{table}
The flavor and CP violations can be in principle independent of each
other in flavored $U(1)'$ models. In particular, the CP symmetry is
respected by the renormalizable scalar potential in Eq.~(\ref{s1}),
thus the model should be extended by the higher-dimensional operators
in~(\ref{s30}) for a nontrivial CP phase beyond the SM\@.
The full scalar potential $V(\phi_i)$ is composed of $V=V_1 + V_2$
with $V_1$ and $V_2$ given in Eqs.~(\ref{s1}) and (\ref{s30}), respectively.

For completeness and the concrete discussion on CP-violating Yukawa couplings to heavy Higgs bosons, we also introduce the $U(1)'$ invariant  Lagrangian for  the renormalizable Yukawa couplings for quarks and leptons,\footnote{
The dimension-5 operators with the singlet $S'$ for the third generation quarks are also consistent with the $U(1)'$ symmetry, as follows,
\begin{equation*}
\Delta {\cal L}_Y= \frac{c'_t}{\Lambda}\, S^{\prime\dagger} {\tilde H}'_2 {\bar q}'_3 u'_3 + \frac{c'_b}{\Lambda} \, S' H'_2 {\bar q}'_3 d'_3 +{\rm h.c.}.
\end{equation*}
Then, there could appear extra CP phases from $c'_t$ and $c'_b$, but we regard them to be model-dependent and ignored in the later discussion.
} which is given by
\begin{align}
  -{\cal L}_Y=
  &~{\bar q}^\prime_i ( y^{\prime u}_{ij}{\tilde H}'_1+ h^{\prime u}_{ij}{\tilde H}'_2 ) u'_j
    +{\bar q}^\prime_i ( y^{\prime d}_{ij} {H}'_1+h^{\prime d}_{ij} {H}'_2 ) d'_j  \nonumber \\
  &+y^{\prime \ell}_{ij} {\bar \ell}^\prime_i {H}'_1 e'_j + y^{\prime\nu}_{ij} {\bar \ell}^\prime_i
    {\tilde H}'_1 \nu^\prime_{jR}
    + \mathrm{h.c.}
\end{align}
with ${\tilde H}'_{1,2}\equiv i\sigma_2 H^{\prime *}_{1,2}$.
The model further includes three right-handed neutrinos and additional
singlet scalars necessary for neutrino masses and mixings.
Here we have left out the additional fields since they are mostly
irrelevant to our study.

As a result, we can fix the flavor structure in the quark
and lepton sectors in the presence of the flavored $U(1)'$ and
investigate the effect of the model-independent dimension-6
operator, $S^{\prime 2} {(H^{\prime\dagger}_1 H'_2)}^2$, for the CP-violating observables such as eEDM\@.
Our later discussion applies to the
general 2HDMs with $U(1)'$ beyond any flavor-specific $U(1)'$, as far
as there is no significant violation of flavor in the Yukawa
couplings for charged leptons.

\section{CP violation in the Higgs sector}

Considering the higher-dimensional terms in the effective scalar
potential of the benchmark models with flavored $U(1)'$, we discuss
the Higgs spectrum and the mixings among CP-even and CP-odd scalars.

\subsection{Scalar mass matrix with CP violation}

In unitary gauge, the Higgs doublet and singlet fields including the
CP phases can be expressed by
\begin{align}
  H'_j &= e^{i\theta_j}\begin{pmatrix}
    \phi^+_j \\
    (v_j+\rho_j+i\eta_j)/\sqrt{2}
  \end{pmatrix}, \quad (j = 1, \, 2),   \nonumber\\
  S'&=\frac{1}{\sqrt{2}}\,e^{i\theta_S}\left(v_s+S_R+iS_I\right) .
\end{align}
We can always make the phase rotations to make all the vacuum
expectation values (VEVs) real as follows:
\begin{align}
H_j&=   e^{-i\theta_j}\, H'_j,  \label{ph1} \\
S&= e^{-i\theta_S}\, S' .  \label{ph2}
\end{align}
Then, the scalar potential terms take the same forms as the ones given
in Eqs.~(\ref{s1}) and~(\ref{s30})
with the complex parameters being redefined by
\begin{align}
\mu&= e^{i(\theta_2-\theta_1+\theta_S)}\, \mu',  \label{para1} \\
c_1 &=  e^{2i(\theta_2-\theta_1+\theta_S)}\,c'_1. \label{para4}
\end{align}
We note that the real parameters in the potential are unchanged under
the phase rotations, so we have changed the notations from primed to
unprimed, {\em e,g.} from $\mu^\prime$ to $\mu$, etc.
From the tadpole conditions given in Appendix~\ref{app:higgs_sector},
we have a CP phase in the $\mu$ term, supported by the dimension-6
operator of $c_1$.
In comparison, for two Higgs doublets without $U(1)'$, there are extra
CP phases from extra quartic couplings for two Higgs doublets, which
would give rise to a nontrivial CP phase of the $\mu$ term by the
tadpole conditions~\cite{CPansatz}.

The nonzero CP phases come from $\mu=\mu_R+i\mu_I$ and $c_1=c_R+ i c_I$.
Then, in the basis where the CP phases of the VEVs are absorbed into
the complex parameters according to Eqs.~(\ref{ph1}) and~(\ref{ph1}),
we obtain  the squared mass matrix for neutral scalar fields with
$(\rho_1,\rho_2,S_R, \eta_1,\eta_2,S_I)$, given by
\begin{equation}
M^2= \left(\begin{array}{cc} M^2_S & M^2_{\rm mix} \\  (M^2_{\rm mix})^T  & M^2_P  \end{array} \right) \label{higgsmass}
\end{equation}
where
the mass matrices for CP-even, CP-odd scalars and the mixing mass matrix are
\begin{equation}
  M^2_S =
  {\tiny \begin{pmatrix}
    2 \lambda_1 v_1^2+\frac{\mu_R v_2 v_s}{\sqrt{2} v_1}&2 v_1 v_2 (\lambda_3+\lambda_4)-\frac{\mu_R v_s}{\sqrt{2}}+\frac{c_R}{\Lambda^2}\,v_1v_2 v_s^2 & 2\kappa_1 v_1 v_s-\frac{\mu_R v_2}{\sqrt{2}} +\frac{c_R}{\Lambda^2}\,v_1 v_2^2 v_s\vspace{0.2cm} \\
    2 v_1 v_2 (\lambda_3+\lambda_4)-\frac{\mu_R v_s}{\sqrt{2}}+\frac{c_R}{\Lambda^2}\,v_1 v_2^2 v_s & 2 \lambda_2 v_2^2+\frac{\mu_R v_1 v_s}{\sqrt{2}v_2} & 2\kappa_2 v_2 v_s-\frac{\mu_R v_1}{\sqrt{2}}+\frac{c_R}{\Lambda^2}\,v^2_1 v_2 v_s\vspace{0.2cm} \\
    2 \kappa_1 v_1  v_s-\frac{\mu_R v_2}{\sqrt{2}}+\frac{c_R}{\Lambda^2}\,v_1 v^2_2 v_s & 2  \kappa_2 v_2  v_s-\frac{\mu_R v_1}{\sqrt{2}}+\frac{c_R}{\Lambda^2}\,v^2_1 v_2 v_s & 2\lambda_S  v_s^2+\frac{\mu_R  v_1 v_2}{\sqrt{2} v_s}
  \end{pmatrix}, } \label{h0matrix}
  \end{equation}
  \begin{equation}
    M^2_P =
    \begin{pmatrix}
      \frac{\mu_R v_2 v_s}{\sqrt{2}v_1}-\frac{c_R}{\Lambda^2}\,v^2_2 v^2_s  & -\frac{1}{\sqrt{2}}\mu_R\, v_s+\frac{c_R}{\Lambda^2}\,v_1v_2 v^2_s & -\frac{1}{\sqrt{2}}\mu_R\, v_2+\frac{c_R}{\Lambda^2}\,v_1v^2_2 v_s  \vspace{0.2cm}\\
      -\frac{1}{\sqrt{2}}\mu_R\, v_s+\frac{c_R}{\Lambda^2}\,v_1v_2 v^2_s  &   \frac{\mu_R v_1 v_s}{\sqrt{2}v_2}-\frac{c_R}{\Lambda^2}\,v^2_1 v^2_s  &  \frac{1}{\sqrt{2}}\mu_R\, v_1-\frac{c_R}{\Lambda^2}\,v^2_1v_2 v_s  \vspace{0.2cm}\\
      -\frac{1}{\sqrt{2}}\mu_R\, v_2+\frac{c_R}{\Lambda^2}\,v_1v^2_2 v_s   &  \frac{1}{\sqrt{2}}\mu_R\, v_1-\frac{c_R}{\Lambda^2}\,v^2_1v_2 v_s  &  \frac{\mu_R v_1 v_2}{\sqrt{2}v_s}-\frac{c_R}{\Lambda^2}\,v^2_1 v^2_2
    \end{pmatrix},  \label{a0matrix}
  \end{equation}
  and
  \begin{equation}
    M^2_{\rm mix} = \frac{1}{\sqrt{2}}\mu_I
    \begin{pmatrix}
      \frac{v_2 v_s}{v_1 } &- v_s & -  v_2 \vspace{0.2cm} \\
      v_s  &   - \frac{v_1v_s}{ v_2}  &  - v_1 \vspace{0.2cm} \\
      v_2  &  -v_1  &   - \frac{v_1 v_2}{v_s}
    \end{pmatrix}.  \label{mix}
  \end{equation}
Here, we have used the tadpole condition, which gives rise to the relation between the CP-violating mass term and the coefficient of the dimension-6 operator,
\begin{equation}
  \frac{c_I}{\Lambda^2} =\frac{\sqrt{2} \mu_I}{v_1 v_2 v_s}. \label{tadpole}
\end{equation}
The detailed discussion on the minimization of the potential and the tadpole conditions can be referred to Appendix~\ref{app:higgs_sector}, upon which the above result is based.

The imaginary part of the $\mu$ parameter, $\mu_I$, mixes the CP-even scalars and the CP-odd scalar.
We note that the original basis is related to the new basis by
\begin{equation}
  \begin{pmatrix}
    \rho_1 \\ \rho_2 \\ S_R \\ \eta_1 \\ \eta_2 \\ S_I
  \end{pmatrix} =
  \mathcal{R}
  \begin{pmatrix}
    \rho_1 \\ \rho_2 \\ S_R \\ A^0 \\ G_Y \\ G^\prime
  \end{pmatrix} ,\label{fullrot}
\end{equation}
where $G_Y$ and $G^\prime$ are the would-be Goldstone bosons
for the spontaneously broken $U(1)_Y \times U(1)'$.
See Appendix~\ref{app:higgs_mixing} for the detailed expressions for
the scalar fields.
The $6 \times 6$ rotation matrix is given as
\begin{equation}
  \mathcal{R} =
  \begin{pmatrix}
    1_{3\times 3} & 0 \\
    0 & \mathcal{R}_3
  \end{pmatrix} ,
\end{equation}
where ${\cal R}_{3}$ is given in~(\ref{r3}).

\subsection{Mixing between CP-even and odd scalars}

Using the results in the previous subsection and choosing a new basis diagonalizing the $3\times 3$ sub-matrix for CP-even scalars in Appendix~\ref{app:higgs_mixing}, the  $4\times 4$ matrix in Eq.~(\ref{h1matrix}) becomes
\begin{equation}
  R_h M^2_{4\times 4}R^T_h= \begin{pmatrix} R_h M^2_{3\times 3} R^T_h & R_h E \\
E^T R^T_h &  m^2_{h^0_4}
    \end{pmatrix}
\end{equation}
with
\begin{equation}
E=\frac{\mu_I v_s  }{\sqrt{2}} N^{-1}_A  \begin{pmatrix}
  \frac{v}{v_1}\vspace{0.2cm} \\   \frac{v}{v_2} \vspace{0.2cm} \\
  \frac{v}{v_s} \end{pmatrix}, \qquad N_A=\frac{1}{\sqrt{1+\frac{v^2_1
      v^2_2}{v^2v^2_s}}}\,.
\label{eq:NA}
\end{equation}
Treating the off-diagonal entries as perturbations, we obtain the approximate mass eigenvalues and mass eigenstates as follows.
\begin{equation}
m^2_{h_n}= m^2_{h^0_n}  + \Delta_{nn} +\sum_{k\neq n} \frac{|\Delta_{nk}|^2}{m^2_{h^0_n}-m^2_{h^0_k}}+\cdots,
\end{equation}
where $m^2_{h_n}$ are the mass eigenvalues for zero off-diagonal components containing the fourth row or column~\cite{Bian:2017xzg} up to the corrections from the dimension-6 operator, given in Eqs.~(\ref{h0s}) and (\ref{h0a}) in Appendix~\ref{app:higgs_mixing},
and
\begin{equation}
h_n= h^0_n +\sum_{k\neq n} \frac{\Delta_{kn}}{m^2_{h^0_n}-m^2_{h^0_k}}\, h^0_k+\cdots
\end{equation}
with
\begin{equation}
\Delta=  \begin{pmatrix} 0 & R_h E \\
E^T R^T_h &  0
    \end{pmatrix}.
\end{equation}

\begin{figure}[tbp]
  \centering
  \includegraphics[width=0.45\textwidth]{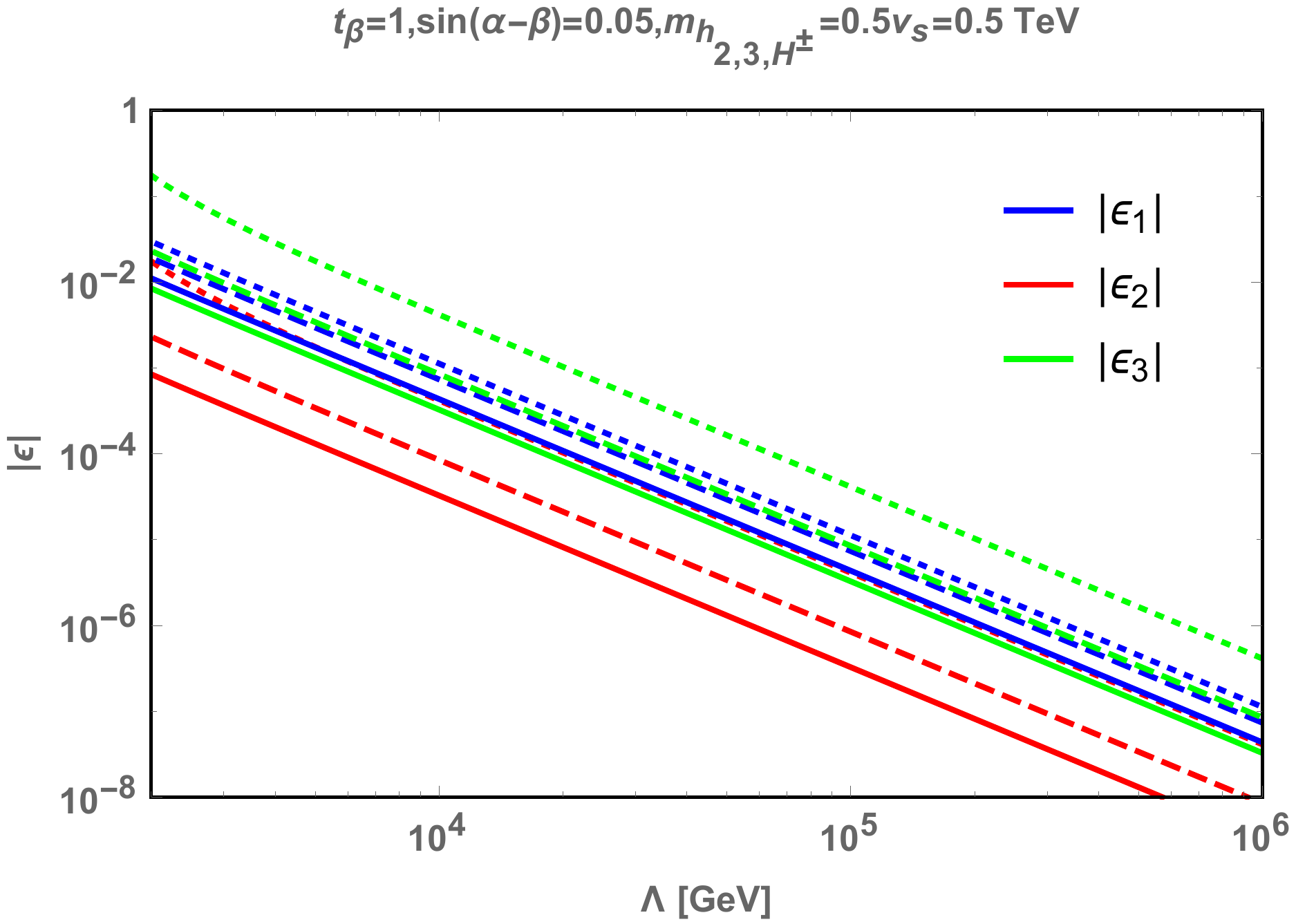} \,\,
  \includegraphics[width=0.45\textwidth]{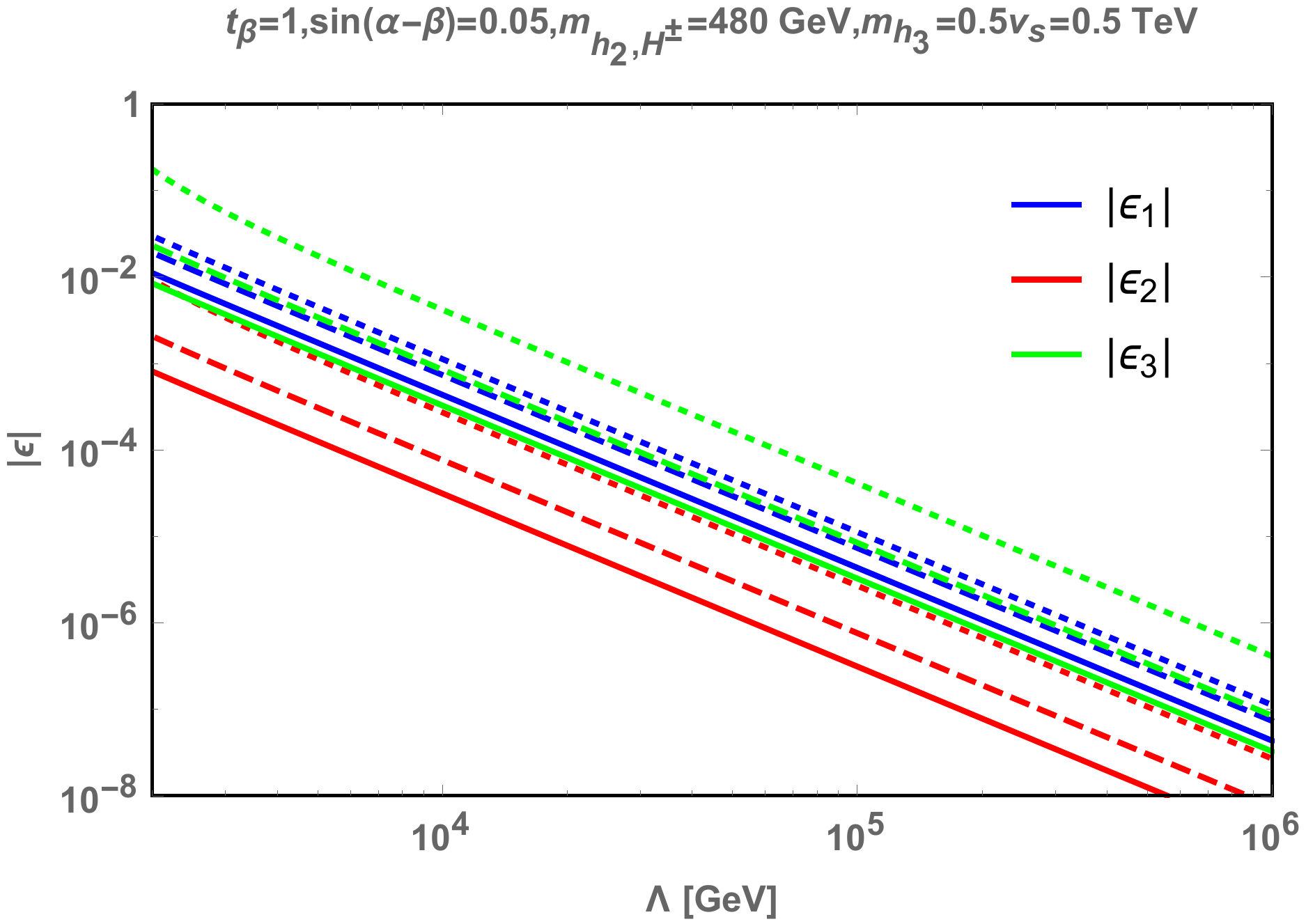} \\
  \caption{The CP-violating mixing parameters, $\varepsilon_i$ ($i=1$,
    $2$, $3$), as a function of the cutoff scale for degenerate and
    non-degenerate case, see Fig.~\ref{fig:de} for eEDM
    predictions. The dotted, dashed, and solid lines represent the
    case for $\mu_R=200$, $300$, and $500$~GeV, respectively.
    We have taken $\tan\beta = 1$, $\sin(\alpha - \beta) = 0.05$, and
    $v_s = 1$~TeV.}
  \label{fig:eps}
\end{figure}

When the Higgs mixings with the CP-even singlet scalar are small, the rotation matrix among CP-even scalars is approximated as\footnote{We note that our conventions for the Higgs mixing are different from those in the main text of our previous work~\cite{Bian:2017xzg}.}
\begin{equation}
  R_h \simeq
  \begin{pmatrix}
    \cos\alpha & \sin\alpha & 0 \\
    -\sin\alpha    & \cos\alpha
    & 0 \\
    0 & 0 & 1
  \end{pmatrix}
\end{equation}
with $\alpha=\alpha_1$ and $\alpha_2\simeq \alpha_3\simeq 0$.
In this case, we can further simplify the above results as
\begin{align}
m^2_{h_1} &\approx m^2_{h^0_1} + \varepsilon_1 (R_h E)_1, \\
m^2_{h_2} &\approx   m^2_{h^0_2} +  \varepsilon_2 (R_h E)_2, \\
m^2_{h_3} &\approx  m^2_{h^0_3} +  \varepsilon_3 (R_h E)_3, \\
m^2_{h_4} &\approx m^2_{h^0_4} - \varepsilon_1 (R_h E)_1  - \varepsilon_2 (R_h E)_2 -  \varepsilon_3 (R_h E)_3,
\end{align}
and
\begin{align}
h_1 &\approx c_\alpha\, \rho_1+ s_\alpha\, \rho_2 +  \varepsilon_1\, A^0, \\
h_2 &\approx  -s_\alpha\, \rho_1+c_\alpha\, \rho_2 +   \varepsilon_2\, A^0, \\
h_3 &\approx S_R +  \varepsilon_3\, A^0, \\
h_4 &\approx A^0+ (- c_\alpha\varepsilon_1 +s_\alpha\varepsilon_2 )\, \rho_1- (s_\alpha\varepsilon_1+c_\alpha\varepsilon_2)\, \rho_2- \varepsilon_3\, S_R
\end{align}
We have used the shorthand notation: $s_\alpha \equiv \sin\alpha$ and
$c_\alpha \equiv \cos\alpha$.
Noting that
$\varepsilon_n \equiv (R_h E)_{n}/(m^2_{h^0_n}-m^2_{h^0_4})$,
which results in
\begin{align}
\varepsilon_1 &= \frac{1}{m^2_{h^0_1}-m^2_{h^0_4}} \frac{\mu_I v_s}{\sqrt{2} N_A}\, \Big(\frac{s_\alpha}{s_\beta} +\frac{c_\alpha}{c_\beta}  \Big), \label{ep1} \\
\varepsilon_2 &= \frac{1}{m^2_{h^0_2}-m^2_{h^0_4}} \frac{\mu_I v_s}{\sqrt{2} N_A}\, \Big(\frac{c_\alpha}{s_\beta} -\frac{s_\alpha}{c_\beta}  \Big),  \label{ep2} \\
\varepsilon_3 &= \frac{1}{m^2_{h^0_3}-m^2_{h^0_4}} \frac{\mu_I v_s}{\sqrt{2} N_A},\label{ep3}
\end{align}
we can write the original scalar fields in terms of the approximate mass eigenstates as
\begin{align}
 \rho_1 &\approx c_\alpha\, h_1-s_\alpha\, h_2 +  (-c_\alpha \varepsilon_1+s_\alpha \varepsilon_2)\, h_4,  \label{rs1} \\
\rho_2 &\approx s_\alpha\, h_1+c_\alpha\, h_2 +   (-s_\alpha \varepsilon_1-c_\alpha \varepsilon_2)\, h_4, \label{rs2}  \\
S_R &\approx h_3 -  \varepsilon_3\, h_4,  \label{rs3} \\
A^0 &\approx h_4 +\varepsilon_1\, h_1+\varepsilon_2\, h_2+  \varepsilon_3\, h_3.  \label{psa}
\end{align}
As a consequence, we find from Eqs.~(\ref{ep1})--(\ref{ep3}) that
close to the alignment limit, where $\alpha=\beta$ for $\tan\beta \equiv v_1
/ v_2$, the CP-violating parameters in the Higgs mixing become
\begin{align}
\varepsilon_1 &\simeq -1.2\times 10^{-3}\, \bigg( \frac{625\,{\rm GeV}}{m_{h^0_4}+m_{h^0_1}} \bigg)\bigg(\frac{375\,{\rm GeV}}{m_{h^0_4}-m_{h^0_1}}\bigg)\bigg(\frac{v_s}{1\,{\rm TeV}}\bigg)\bigg( \frac{\mu_I}{0.2\,{\rm GeV}}\bigg),\label{epsilon1} \\
\varepsilon_2 &\simeq-4.4\times 10^{-3}\, \bigg( \frac{950\,{\rm GeV}}{m_{h^0_4}+m_{h^0_2}} \bigg)\bigg(\frac{50\,{\rm GeV}}{m_{h^0_4}-m_{h^0_2}}\bigg)\bigg(\frac{v_s}{1\,{\rm TeV}}\bigg)\bigg( \frac{\mu_I}{0.2\,{\rm GeV}}\bigg), \label{epsilon2}\\
\varepsilon_3 &\simeq-2.9\times 10^{-3}\, \bigg( \frac{950\,{\rm GeV}}{m_{h^0_4}+m_{h^0_3}} \bigg)\bigg(\frac{50\,{\rm GeV}}{m_{h^0_4}-m_{h^0_3}}\bigg)\bigg(\frac{v_s}{1\,{\rm TeV}}\bigg)\bigg( \frac{\mu_I}{0.2\,{\rm GeV}}\bigg).\label{epsilon3}
\end{align}
Here, the typical value for $\mu_I$ was taken from Eq.~(\ref{tadpole}) for $\Lambda/\sqrt{|c_I|}\simeq10$~TeV.

In Fig.~\ref{fig:eps}, we depict the CP-violating mixing parameters,
$|\varepsilon_i|$ ($i=1$, $2$, $3$), as a function of the cutoff scale for different $\mu_R$.
For $\tan\beta=1$ and $\mu_R=100\,{\rm GeV}$--$1\,{\rm
  TeV}$, the VEV of the singlet scalar field is bounded as
$250\,{\rm GeV}\lesssim v_s\lesssim 2\,{\rm TeV}$~\cite{Bian:2017xzg}
in the alignment limit.  Thus, in order to choose a larger $v_s$ for heavy $Z'$, a smaller $\mu_R$ is favored by unitary.
The figure shows that the smaller $\mu_R$, the larger the CP-violating
mixing predicted, as indicated by
Eqs.~(\ref{epsilon1})--(\ref{epsilon3}).
Moreover, the larger the mass splitting between $h_{1,2,3}$ and $h_4$,
the smaller $\varepsilon_i$ is.

\begin{figure}[tbp]
  \centering
  \includegraphics[width=0.48\textwidth]{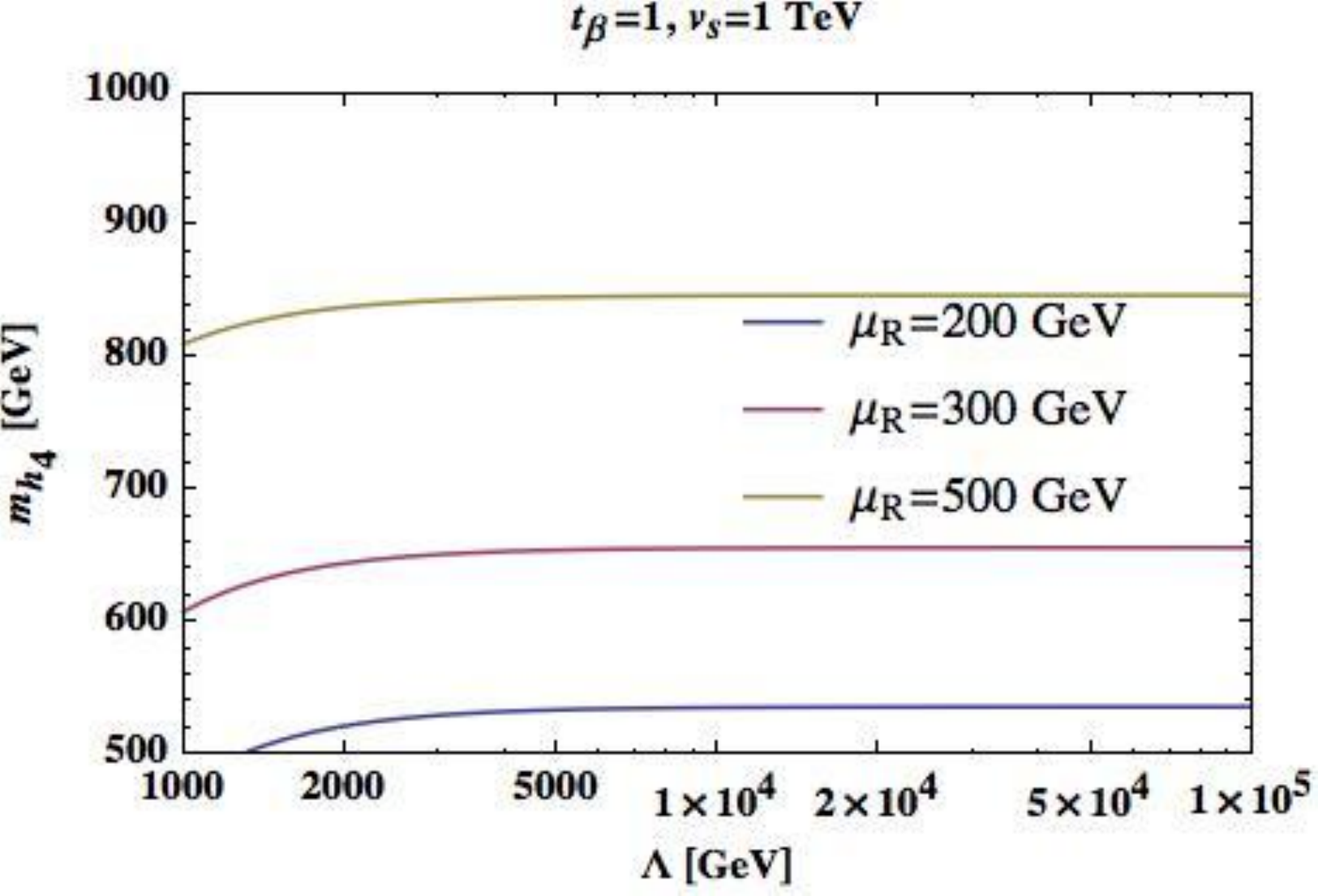}
  \caption{\label{fig:mh4} The mass of the pseudoscalar-like Higgs,
    $m_{h_4}$, {\em vs}.\ the cutoff scale, for $\mu_R=200$, $300$,
    $500$~GeV for upper, middle, and lower lines, respectively.}
\end{figure}

Since the pseudoscalar-like Higgs mixes with the CP-even scalars due to the CP violation, it is important to identify the allowed mass range of the pseudoscalar-like Higgs in our model.
In Fig.~\ref{fig:mh4},  we show the range of masses for the
pseudoscalar-like Higgs $m_{h_4}$ and the cutoff scale. As a result, $m_{h_4}$ becomes almost independent of the cutoff scale with $\Lambda\gtrsim 2\,{\rm TeV}$, so it can be determined mainly by the $\mu_R$ parameter. Thus, we find that as $\mu_R$ increases, $m_{h_4}$ becomes larger, according to Eq.~(\ref{h0a}).

We also remark that the charged Higgs mass is not affected by the CP violation. It is given in Eq.~(\ref{H+mass}) as in Ref.~\cite{Bian:2017xzg}.

\section{Yukawa couplings with CP violation}

In this section,  we present the Yukawa couplings for the SM fermions and new flavor-changing couplings for quarks in our model from the results in the Appendix~\ref{app:quark_diagonal}.  The results provide a complete basis for studying EDMs, collider searches and magnetic dipole moments of leptons in the next section, together with the Higgs mixing in the previous section.

For the phase-rotated scalar fields with Eqs.~(\ref{ph1})--(\ref{ph2}), we can rewrite the Yukawa couplings for quarks and leptons as follows.
\begin{align}
  -{\cal L}_Y=
  &~{\bar q}^\prime_i ( e^{-i\theta_1}y^{\prime u}_{ij}{\tilde H}_1+ e^{-i\theta_2} h^{\prime u}_{ij}{\tilde H}_2 ) u'_j
    +{\bar q}^\prime_i ( e^{i\theta_1} y^{\prime d}_{ij} {H}_1+e^{i\theta_2} h^{\prime d}_{ij} {H}_2 ) d'_j  \nonumber \\
  &+e^{i\theta_1} y^{\prime \ell}_{ij} {\bar \ell}^\prime_i {H}_1 e'_j + e^{-i\theta_1} y^{\prime\nu}_{ij} {\bar \ell}^\prime_i
    {\tilde H}_1 \nu^\prime_{jR}
    + \mathrm{h.c.}
\end{align}
Then, after the scalars get nonzero VEVs, we obtain the quark and lepton mass
terms as
\begin{equation}
  {\cal L}_Y=- {\bar u}'_L M_u u'_R-{\bar d}'_L M_d d'_R - {\bar \ell}'_L M_\ell' \ell'_R
  - {\bar \ell}'_L M_D \nu'_R
  + \mathrm{h.c.}
\end{equation}
with the following flavor structure:
\begin{align}
  M_u
  &= \begin{pmatrix}
    y^u_{11}\langle  {\tilde H}_1\rangle &
   y^u_{12}\langle {\tilde H}_1\rangle & 0 \\
    y^u_{21} \langle {\tilde H}_1\rangle &
    y^u_{22} \langle {\tilde H}_1 \rangle &  0 \\
    h^u_{31} \langle {\tilde H}_2 \rangle &
     h^u_{32}\langle {\tilde H}_2\rangle &
    y^u_{33} \langle {\tilde H}_1 \rangle
  \end{pmatrix},\label{qmass1} \\
  M_d
  &= \begin{pmatrix}
    y^d_{11}\langle { H}_1\rangle &
   y^d_{12}\langle { H}_1\rangle &
    h^d_{13} \langle {H}_2\rangle \\
  y^d_{21} \langle { H}_1 \rangle &
    y^d_{22} \langle {H}_1\rangle &
    h^d_{23}\langle {H}_2\rangle \\
    0 &  0 &   y^d_{33}  \langle { H}_1 \rangle
  \end{pmatrix},\label{qmass2} \\
  M_\ell
  &=  \begin{pmatrix}
   y^\ell_{11} \langle { H}_1 \rangle & 0 & 0 \\
    0 & y^\ell_{22}  \langle { H}_1 \rangle & 0 \\
    0 & 0 & y^\ell_{33} \langle { H}_1\rangle
    \end{pmatrix},\label{cleptonmass} \\
  M_D
  &= \begin{pmatrix}
    y^\nu_{11} \langle {\tilde H}_1 \rangle & 0 & 0 \\
    0 & y^\nu_{22}\langle {\tilde H}_1 \rangle & 0 \\
    0 & 0 &  y^\nu_{33} \langle {\tilde H}_1 \rangle
  \end{pmatrix}.
\end{align}
Here, we absorbed the Higgs phases into the Yukawa couplings by $y^u_{ij}=e^{-i\theta_1} y^{\prime u}_{ij}$, $h^u_{ij}=e^{-i\theta_2} h^{\prime u}_{ij}$, etc.
Since the mass matrix for charged leptons is already diagonal, the
lepton mixings come from the mass matrix of right-handed neutrinos.

\subsection{Quark Yukawa couplings}

We begin with the quark Yukawa couplings to the neutral scalars in the interaction basis,
\begin{align}
 -\mathcal{L}^h_{Y}=&~ {\bar d}_L \bigg[\frac{1}{v_1}\,M^D_d (\rho_1+i\eta_1) +\frac{1}{\sqrt{2}}\Big(-\frac{v_2}{v_1} (\rho_1+i\eta_1)+(\rho_2+i\eta_2) \Big) {\tilde h}^d \bigg] d_R \nonumber \\
 & +{\bar u}_L \bigg[\frac{1}{v_1}\,M^D_u (\rho_1-i\eta_1) +\frac{1}{\sqrt{2}}\Big(-\frac{v_2}{v_1} (\rho_1-i\eta_1)+(\rho_2-i\eta_2) \Big) {\tilde h}^u \bigg] u_R  +{\rm h.c}.
\end{align}
Then, using the Higgs mixing Eqs.~(\ref{rs1})--(\ref{psa}) in the previous section and Eqs.~(\ref{ps1})--(\ref{ps3}),
the Yukawa terms for the third-generation quarks are now written
as
\begin{align}
  -\mathcal{L}^h_Y \supset
  &~ \frac{1}{\sqrt{2}}
    \sum_{i = 1}^4 \left[
    \left( \lambda_t^{h_i} + i \widetilde\lambda_t^{h_i}
    \right) \bar t_L t_R h_i
    + \left( \lambda_b^{h_i} + i \widetilde\lambda_b^{h_i}
    \right) \bar b_L b_R h_i \right] \nonumber\\
  & - \frac{s_{\beta - \alpha}}{\sqrt{2} c_\beta}
    \bar b_L (\widetilde{h}_{13}^d d_R + \widetilde{h}_{23}^d s_R ) h_1
    + \frac{c_{\beta - \alpha}}{\sqrt{2} c_\beta}
    \bar b_L (\widetilde{h}_{13}^d d_R + \widetilde{h}_{23}^d s_R )
    h_2 \nonumber\\
  & - \frac{i N_A}{\sqrt{2} c_\beta} \bar b_L \left(
    \widetilde{h}_{13}^d d_R + \widetilde{h}_{23}^d s_R \right)
    ( \varepsilon_1 h_1 + \varepsilon_2 h_2 + \varepsilon_3 h_3 + h_4)
    + \mathrm{h.c.}, \label{FCYukawas}
\end{align}
where
\begin{align}
  \lambda_{t}^{h_1}
  &= \frac{\sqrt{2} m_{t} c_\alpha}{v c_\beta} ,
    \quad
  \lambda_{t}^{h_2}
  = - \frac{\sqrt{2} m_{t} s_\alpha}{v c_\beta} ,
    \quad
  \lambda_{t}^{h_3}
  = 0, \quad
  \lambda_{t}^{h_4}
  =  \frac{\sqrt{2} (- c_\alpha \varepsilon_1 + s_\alpha
    \varepsilon_2) m_{t}}{v c_\beta}
    \nonumber\\
  \lambda_{b}^{h_1}
  &= \frac{\sqrt{2} m_{b} c_\alpha}{v c_\beta} -
    \frac{\widetilde h_{33}^d s_{\beta - \alpha}}{c_\beta} ,
    \quad
  \lambda_{b}^{h_2}
    = - \frac{\sqrt{2} m_{b} s_\alpha}{v c_\beta}
    + \frac{\widetilde h_{33}^d c_{\beta - \alpha}}{c_\beta} ,
    \quad
  \lambda_{b}^{h_3}
  = 0, \nonumber\\
  \lambda_{b}^{h_4}
  &=  \frac{\sqrt{2} (- c_\alpha \varepsilon_1 + s_\alpha
    \varepsilon_2) m_{b}}{v c_\beta}
    + \frac{\widetilde h_{33}^d (s_{\beta - \alpha} \varepsilon_1 -
    c_{\beta - \alpha} \varepsilon_2)}{c_\beta} ,
    \nonumber\\
  \widetilde\lambda_t^{h_i}
  &= - N_A \varepsilon_i \frac{\sqrt{2} m_t t_\beta}{v}
    \quad (i = 1, \, 2, \, 3) ,
    \quad
  \widetilde\lambda_t^{h_4}
  = - N_A \frac{\sqrt{2} m_t t_\beta}{v} ,
    \nonumber\\
  \widetilde\lambda_b^{h_i}
  &= N_A \varepsilon_i \left( \frac{\sqrt{2} m_b t_\beta}{v} -
    \frac{\widetilde{h}_{33}^d}{c_\beta }\right)
    \quad (i = 1, \, 2, \, 3) ,
    \quad
  \widetilde\lambda_b^{h_4}
  = N_A \left( \frac{\sqrt{2} m_b t_\beta}{v} -
    \frac{\widetilde{h}_{33}^d}{c_\beta }\right) .
\end{align}
Here, ${\tilde h}^d\equiv  D^\dagger_L h^d D_R$ and
${\tilde h}^u\equiv   U^\dagger_L h^u U_R$. Thus, by taking $U_L=1$ we
get ${\tilde h}^u=h^u U_R$ and ${\tilde h}^d=V^\dagger_{\rm CKM} h^d$.
We note that $\lambda_t^{h_3}$ and $\lambda_b^{h_3}$ are vanishing because we
have neglected the mixing.

As compared to the type-I 2HDM, we have extra Yukawa couplings given by
\begin{align}
  {\tilde h}^d_{13}
  &= 1.80\times 10^{-2}\Big(\frac{m_b}{v\sin\beta}\Big),  \label{ht13} \\
  {\tilde h}^d_{23}
  &=5.77\times 10^{-2}\Big(\frac{m_b}{v\sin\beta}\Big) , \label{ht23} \\
  {\tilde h}^d_{33}
  &= 2.41\times 10^{-3}\Big(\frac{m_b}{v\sin\beta}\Big).  \label{ht33}
\end{align}
We find that there is no modification in the top-quark Yukawa coupling as compared to the SM, whereas down-type quarks can have large flavor-violating couplings  if $\tan\beta$ is small. In the alignment limit where $\alpha=\beta$, the flavor-violating interactions of the SM-like Higgs $h_1$
boson are turned off.
In Refs.~\cite{Bian:2017rpg,Bian:2017xzg},
we have discussed the phenomenological bounds on
the sizable flavor-changing couplings for down-type quarks, for
instance, the bounds from $B$-meson decays ($B_s\rightarrow
\mu^+\mu^-$, $B_s\rightarrow X\gamma $) and mixings ($B_s$--${\bar
  B}_s$), etc, constrain the parameter space for heavy Higgs scalar
and $Z'$ masses.

From the resulting Yukawa couplings in~(\ref{FCYukawas}), the simultaneous presence of scalar and pseudoscalar couplings violate the CP symmetry. In particular, the CP-violating top Yukawa couplings are constrained by the bounds from neutron and electron EDMs.
Note that in the alignment limit where $\alpha = \beta$ and
$\varepsilon_i \ll 1$ {\em i.e.}, $\mu_I$, $m_{h_1}$, $m_{h_2}$, $m_{h_3} \ll
m_{h_4}$, the CP violation arises mainly through $h_2$ and $h_4$.
There are also usual flavor-diagonal Yukawa couplings of neutral scalars to light quarks, including the CP-violating mixing, but they are sub-dominant for the EDM contributions.

The Yukawa terms of the charged Higgs boson are given as
\begin{equation}
  -\mathcal{L}_{Y}^{H^-} =
  {\bar b}(\lambda_{t_L}^{H^-} P_L + \lambda_{t_R}^{H^-} P_R) t H^-
  + {\bar b}(\lambda_{c_L}^{H^-} P_L + \lambda_{c_R}^{H^-} P_R ) c H^-+
    \lambda_{u_L}^{H^-} {\bar b}P_L u H^- + \mathrm{h.c.},
\end{equation}
where
\begin{align}
  \lambda_{t_L}^{H^-}
  &= \frac{\sqrt{2}m_b \tan\beta}{v}\, V^*_{tb}
    - \frac{(V_{\rm CKM} {\tilde h}^d)^*_{33}}{\cos\beta},
  \nonumber\\
  \lambda_{t_R}^{H^-}
  &= -\frac{\sqrt{2} m_t \tan\beta}{v}
    \,V^*_{tb},
  \nonumber\\
  \lambda_{c_L}^{H^-}
  &= \frac{\sqrt{2}m_b \tan\beta}{v}\, V^*_{cb}
    -  \frac{(V_{\rm CKM}{\tilde h}^d)^*_{23}}{\cos\beta},
  \nonumber\\
  \lambda_{c_R}^{H^-}
  &= -\frac{\sqrt{2} m_c\tan\beta}{v}\, V^*_{cb},
  \nonumber\\
  \lambda_{u_L}^{H^-}
  &= \frac{\sqrt{2}m_b \tan\beta}{v}\, V^*_{ub}
    -  \frac{(V_{\rm CKM}{\tilde h}^d)^*_{13}}{\cos\beta}
\end{align}
with
\begin{equation}
  V_{\rm CKM}{\tilde h}^d=
  \begin{pmatrix}
    0 & 0 & V_{ud}{\tilde h}^d_{13}+ V_{us}{\tilde h}^d_{23}+V_{ub}{\tilde h}^d_{33} \\
    0 & 0 & V_{cd}{\tilde h}^d_{13}+ V_{cs}{\tilde h}^d_{23}+V_{cb}{\tilde h}^d_{33} \\
    0 & 0 & V_{td}{\tilde h}^d_{13}+ V_{ts}{\tilde h}^d_{23}+V_{tb}{\tilde h}^d_{33}
  \end{pmatrix}.
\end{equation}

\subsection{Lepton Yukawa couplings}

As can be seen in~(\ref{cleptonmass}), the mass matrix for charged leptons
$e_j$ is already diagonal due to the $U(1)'$ symmetry. Thus, the
lepton Yukawa couplings are in a flavor-diagonal form, given by
\begin{align}
  -{\cal L}_{Y}^\ell
  =&~\frac{m_{e_j}\cos\alpha }{v\cos\beta}\, {\bar e}_j\, e_j \,h_1
   - \frac{m_{e_j}\sin\alpha }{v\cos\beta} \,{\bar e}_j\, e_j \,h_2 + \frac{m_{e_j} }{v\cos\beta} \,{\bar e}_j\, e_j \,(-c_\alpha\varepsilon_1+s_\alpha \varepsilon_2)h_4 \nonumber \\
   & +\frac{i m_{e_j}\, N_A \tan\beta}{v}\, {\bar e}_j \gamma^5 e_j \,(h_4+ \varepsilon_1\, h_1+ \varepsilon_2\, h_2+  \varepsilon_3\, h_3)
    \nonumber\\
  &+\frac{\sqrt{2}m_{e_j}\tan\beta}{v}\, \left({\bar \nu}_j\,P_R\,
    e_j \,H^+ + \mathrm{h.c.}\right).
\end{align}
As a result, the CP symmetry is also broken in the lepton Yukawa couplings to the neutral scalars.

\section{\label{sec:EDM_collider}\boldmath{$B$}-meson anomalies, EDM and collider searches}

We update the status of $B$-meson anomalies in light of the updated
data and analysis on $R_{K^{(*)}}$ ratios and review the parameter
space for $Z'$ mass and couplings for the flavored $U(1)'$
model. Then, we calculate the EDM of the electron in the presence of the CP-violating mixings between neutral scalars and constrain the extra Higgs masses and the cutoff scale for higher-dimensional operators. We also briefly discuss the anomalous magnetic moments of leptons in our model.

\subsection{Bounds from \boldmath{$B$}-meson decays}

We first remark that the measurement of $R_K={\cal B}(B\rightarrow K\mu^+\mu^-)/{\cal
  B}(B\rightarrow Ke^+e^-)$ has been updated by the new analysis with LHCb 2015--2016 data~\cite{RK-new}, showing the combined value with LHCb 2011--2012 data,
\begin{equation}
R_K=0.846^{+0.060}_{-0.054}({\rm stat})^{+0.016}_{-0.014}({\rm syst}),
\end{equation}
which deviates from the SM prediction by $2.5\sigma$.\footnote{
  After the completion of our work, the LHCb collaboration has announced the updated
  result on the $R_K$ variable for $B$-meson decays using the integrated
  luminosity of 9~fb$^{-1}$ in Run 1 and Run 2~\cite{Aaij:2021vac},
  which is $R_K=0.846^{+0.042}_{-0.039}({\rm
    stat})^{+0.013}_{-0.012}({\rm syst})$. The central value remains
  the same, but the uncertainties have been reduced. Thus, the
  deviation from the SM prediction in $R_K$ is now 3.1$\sigma$, which
  would hint at the violation of lepton universality.
  The following discussion on the $B$-meson anomalies is qualitatively
  intact under the updated result.
}
The updated global fit for $B$-meson decays shows that the purely muonic contribution from new physics to the Wilson coefficients, $C^{\mu,{\rm NP}}_9=-C^{\mu,{\rm NP}}_{10}$, is favored from the data for lepton flavor non-universality~\cite{RK-newfit,update}, as compared to $C_9$ only, but $C^{\mu,{\rm NP}}_9\neq 0$ and $C^{\mu,{\rm NP}}_{10}=0$ is slightly favored for all the data set~\cite{update}.
Recently, the analysis of the full Belle data sample has led to new results on $R_K$ in various bin energies, in particular, the new Belle result in the bin of interest, $1\,{\rm GeV}^2<q^2 <6\,{\rm GeV}^2$, is consistent with the LHCb result~\cite{RK-belle-new}.

For vector $B$-mesons, $R_{K^*}={\cal
  B}(B\rightarrow K^*\mu^+\mu^-)/{\cal B}(B\rightarrow
K^*e^+e^-)$ from LHCb~\cite{RKs} is
\begin{equation}
  R_{K^*}= \left\{ \begin{array}{cc} 0.66^{+0.11}_{-0.07}({\rm stat})\pm 0.03({\rm syst}), \quad 0.045\,{\rm GeV}^2<q^2 <1.1\,{\rm GeV}^2, \vspace{0.2cm} \\  0.69^{+0.11}_{-0.07}({\rm stat})\pm 0.05({\rm syst}), \quad 1.1\,{\rm GeV}^2<q^2 <6.0\,{\rm GeV}^2, \end{array}\right.
\end{equation}
which again differs from the SM prediction by 2.1--$2.3\sigma$ and
2.4--$2.5\sigma$, depending on the energy bins. The deviation in
$R_{K^*}$ is supported by the discrepancy in the angular distribution of
$B\rightarrow K^*\mu^+\mu^-$~\cite{P5} and  the recent update on
$R_{K^*}$ from the Belle data also shows a similar deviation in particular in low energy bins ($0.045\,{\rm GeV}<q^2<1.1\,{\rm GeV}$)~\cite{RKs-new}.

We also remark on the intriguing related anomalies in $R_D={\cal
  B}(B\rightarrow D\tau\nu)/{\cal  B}(B\rightarrow D\ell\nu)$ and
$R_{D^*}={\cal B}(B\rightarrow D^*\tau\nu)/{\cal  B}(B\rightarrow
D^*\ell\nu)$ with $\ell=e$, $\mu$ for BaBar~\cite{babar} and
Belle~\cite{belle,belle-new1}  and $\ell=\mu$ for LHCb~\cite{lhcb}.
In this case, the deviations between the measurements and the
SM predictions for $R_D$ and $R_{D^*}$ are $1.4\sigma$ and
$2.5\sigma$, respectively, amounting to the combined deviation of
$3.08\sigma$~\cite{hflav}. However, the recently measured values of $R_{D^{(*)}}$
with semi-leptonic tagging in Belle~\cite{belle-new} agree with the SM predictions within $1.6\sigma$. $R_{D^{(*)}}$ anomalies are not explained in our model, but they can be easily explained with leptoquarks, also accounting for the anomalous magnetic moment of muon~\cite{LQ}.

After integrating out the $Z'$ gauge boson in our model, we obtain the effective
four-fermion interaction for ${\bar b}\rightarrow {\bar s}\mu^+ \mu^-$
as follows.
\begin{equation}
{\cal L}_{{\rm eff},{\bar b}\rightarrow {\bar s}\mu^+ \mu^-}= -\frac{xy g^2_{Z'}}{3 m^2_{Z'}}\, V^*_{ts} V_{tb}\,  ({\bar s}\gamma^\mu P_L b) ({\bar \mu}\gamma_\mu \mu)+{\rm h.c.}
\end{equation}
Consequently, as compared to the effective Hamiltonian with the SM
normalization,
\begin{equation}
\Delta {\cal H}_{{\rm eff},{\bar b}\rightarrow {\bar s}\mu^+ \mu^-} = -\frac{4G_F}{\sqrt{2}}  \,V^*_{ts} V_{tb}\,\frac{\alpha_\text{em}}{4\pi}\, C^{\mu,{\rm NP}}_9 {\cal O}^\mu_9
\end{equation}
with $ {\cal O}^\mu_9 \equiv ({\bar s}\gamma^\mu P_L b) ({\bar
  \mu}\gamma_\mu \mu)$ and $\alpha_{\rm em}$ being the electromagnetic
coupling, the new physics contribution to the Wilson
coefficient is identified as
\begin{equation}
  C^{\mu, {\rm NP}}_9= -\frac{8 xy \pi^2\alpha_{Z'}}{3\alpha_{\rm em}}\, \left(\frac{v}{m_{Z'}}\right)^2
\end{equation}
with $\alpha_{Z'}\equiv g^2_{Z'}/(4\pi)$,
and vanishing contributions to other operators, $C^{\mu,{\rm
    NP}}_{10}=C^{\prime\mu, {\rm NP}}_9=C^{\prime\mu, {\rm
    NP}}_{10}=0$.
Choosing $xy>0$  for a negative sign of $C^\mu_9$ for $B$-meson anomalies from $R_{K^{(*)}}$
and requiring the best-fit value, $C^{\mu, \, {\rm
    NP}}_9=-0.98$~\cite{update}, (while taking $[-1.15,-0.81]$ and
$[-1.31,-0.64]$ within $1\sigma$ and $2\sigma$ errors), to explain the
$B$-meson anomalies together with the full set of the data~\cite{update}, we get the condition for $Z'$ mass and couplings as follows:
\begin{equation}
  m_{Z'}= 1.27~\text{TeV} \times \left(xy\,
    \frac{\alpha_{Z'}}{\alpha_{\rm em}} \right)^{1/2} .
\end{equation}
Therefore, $m_{Z'} \simeq 1\,{\rm TeV}$ for $xy\simeq 1$ and
$\alpha_{Z'}\simeq \alpha_{\rm em}$. For values of
$xy$ less than unity or $\alpha_{Z'}\lesssim \alpha_{\rm em}$, $Z'$
can be even lighter.

There are various phenomenological constraints on the  $Z'$ for bottom quarks and leptons, coming
from dimuon resonance searches, $B-{\bar B}$ mixings, other meson decays such as $B\rightarrow X_s\gamma $,  tau
lepton decays and neutrino scattering. Taking them into account, it was shown that the parameter space with $x g_{Z'}\lesssim 0.05$ for $y g_{Z'}\simeq 1$ and $m_{Z'}\lesssim 1$~TeV~\cite{Bian:2017rpg,Bian:2017xzg} is consistent for $B$-meson anomalies. See also the phenomenological discussion on similar models in Ref.~\cite{allanach}.

On the other hand, in the presence of sizable flavor violating couplings among down-type quarks and heavy Higgs bosons in Eqs.~(\ref{ht13}) and~(\ref{ht23}), the $B$-meson decays, $B_s\rightarrow \mu^+\mu^-$, $B$--${\bar B}$ mixings, $B\rightarrow X_s\gamma$, etc, can strongly constrain the parameter space for heavy Higgs bosons in combination of unitarity and perturbativity. For instance, in the alignment limit and for $\tan\beta=1$, the masses of heavy Higgs bosons must be in the range of $200\,{\rm GeV}\lesssim m_{h_4}\lesssim 700\,{\rm GeV}$ and $200\,{\rm GeV}\lesssim m_{h_2}=m_{H^+}\lesssim 600\,{\rm GeV}$~\cite{Bian:2017rpg,Bian:2017xzg}.
For a smaller value of $\tan\beta$, all the $B$-meson and theoretical bounds become more stringent, due to larger flavor violating couplings, so the  masses of heavy Higgs bosons should be almost degenerate and about $300$--$400$~GeV.

\subsection{Electric dipole moments}

The current strongest limit on the electron EDM (eEDM) comes from ACMEII~\cite{acme},
 \begin{equation}
 d_e<1.1\times 10^{-29}\,{\rm e\,cm}.
 \end{equation}
In the presence of the mixings among CP-even and CP-odd
scalars, the couplings of physical scalars $h_i$ ($i=1$, $2$, $3$,
$4$) to the SM fermions and the $W$ and $Z$ bosons can be parameterized as
\begin{equation}
\mathcal{L}= \sum_{i=1}^4 \Big[ -m_f\left( c_{f,i} \bar f f+ \tilde
  c_{f,i} \bar f i\gamma_5 f  \right) + a_i \left( 2  m_W^2 W_\mu
  W^\mu +  m_Z^2 Z_\mu Z^\mu \right)  \Big] \frac{h_i}{v}  \,, \label{eqn:couplings}
\end{equation}
with the coefficients $c_{f,i}$, $\tilde{c}_{f,i}$ and $a_i$ shown in the previous section and Appendix~\ref{app:higgs_self}.

We now discuss the contributions of the scalar couplings to the eEDM\@.
For light fermions, the dominant contributions to their eEDM come from the two-loop Barr-Zee type diagrams~\cite{Barr:1990vd}. For the effective operator for the eEDM,
\begin{equation}
{\cal L}_{\rm eff, EDM} = -\frac{i}{2}\,\delta_e\, {\bar e} \sigma_{\mu\nu} \gamma_5 e F^{\mu\nu},
\end{equation}
the Wilson coefficient $\delta_e$ receives various contributions as listed below.
\begin{eqnarray}
\label{eq:eEDMOp_total}
\delta_e =
(\delta_e )_t^{h\gamma\gamma} +
(\delta_e )_t^{h Z \gamma} +
(\delta_e )_W^{h \gamma\gamma} +
(\delta_e )_W^{h Z \gamma} +
(\delta_e )_{H^\pm}^{h \gamma\gamma} +
(\delta_e )_{H^\pm}^{h Z \gamma} +
(\delta_e)_h^{H^\pm W^\mp \gamma} ,
\end{eqnarray}
where the contributions from the diagrams with effective $h_i \gamma\gamma$ and $h_i Z\gamma$ couplings (from integrating out a top quark loop) are, respectively,
\begin{align}
\label{eq:haaOp_top}
\left(\delta_f \right)^{h_i\gamma\gamma}_{t} &= -  \frac{N_c Q_f Q_{t}^2 e^2}{64\pi^4} \sum_{i=1}^4 \Big[
f(z^i_{t}) \, c_{t,i} \tilde c_{f,i} +
g(z^i_{t}) \, \tilde c_{t,i} c_{f,i} \Big] \,, \\
\label{eq:haZOp_top}
\left(\delta_f \right)^{h_i Z\gamma}_{t} &= -  \frac{N_c Q_f g_{Z\bar f f}^V g_{Z\bar tt}^V}{64\pi^4} \sum_{i=1}^4 \left[
\tilde f\left(z^i_{t}, \, \frac{m_{t}^2}{m_Z^2}\right) c_{t,i} \tilde c_{f,i} +
\tilde g\left(z^i_{t}, \, \frac{m_{t}^2}{m_Z^2}\right) \tilde c_{t,i} c_{f,i} \right] \,.
\end{align}
Here $z_X^i \equiv m_X^2 / M_{h_i}^2$, $g_{Z f\bar f}^V$ is the
vector-current couplings of the $Z$ boson to the fermions, and the
loop integral functions are given by
\begin{align}
f(z) &\equiv \frac{z}{2} \int_0^1 dx \frac{1-2x(1-x)}{x(1-x)-z} \ln\frac{x(1-x)}{z} \ , \nonumber\\
g(z) &\equiv \frac{z}{2} \int_0^1 dx \frac{1}{x(1-x)-z} \ln\frac{x(1-x)}{z} \,, \nonumber\\
\tilde f(x\,,y) &\equiv \frac{y f(x) - x f(y) }{y - x } \,,\nonumber\\
\tilde g(x\,,y) &\equiv \frac{y g(x) - x g(y) }{y-x}\,.
\end{align}
The contributions from the $W$ and Goldstone bosons to the
$h_i \gamma\gamma$ and $h_i Z\gamma$ operators are given as
follows~\cite{Chang:1990sf,Leigh:1990kf, Abe:2013qla}:
\begin{align}
\label{eq:haaOp_W}
\left(\delta_f \right)^{h_i \gamma\gamma}_W &=
\frac{Q_f e^2}{256\pi^4} \sum_{i=1}^4 \left[
\Big( 6 + \frac{1}{z^i_W} \Big) f(z^i_w) +
\Big( 10 - \frac{1}{z^i_W} \Big) g(z^i_w)\right.\nonumber\\
&\qquad\qquad\qquad+ \left. \frac{3}{4}
\Big( g(z^i_W) +  h (z^i_W)  \Big)  \right] a_i \tilde c_{f,i}  \ ,  \\
\label{eq:haZOp_W}
\left(\delta_f \right)^{h_i Z\gamma}_W &=  \frac{g_{Z\bar f f}^V g_{ZWW}}{256\pi^4} \sum_{i=1}^4 \left[
\left(6 -\sec^2\theta_W + \frac{2-\sec^2\theta_W}{2z^i_w} \right)\tilde f(z^i_W, c_W^2 ) \right.\nonumber\\
&\qquad\qquad\qquad\qquad+ \left.
\left( 10- 3\sec^2\theta_W - \frac{2-\sec^2\theta_W}{2z^i_w}\right)\tilde g(z^i_W, c_W^2 ) \right. \nonumber\\
&\qquad\qquad\qquad\qquad + \left. \frac{3}{2}
\Big( g( z^i_W ) + h(  z^i_W) \Big)  \right] a_i \tilde c_{f_i}
\end{align}
with the triple gauge coupling $g_{WWZ} = e/\tan\theta_W$.
$h(z)$ is the loop function given by
\begin{eqnarray}
h(z) \equiv \frac{z}{2} \int_0^1 dx \frac{1}{z-x(1-x)} \left( 1+ \frac{z}{z-x(1-x)} \ln\frac{x(1-x)}{z} \right) \,.
\end{eqnarray}
The contributions from the charged Higgs bosons running in loops also read
\begin{align}
\left(\delta_f \right)^{h_i\gamma\gamma}_{H^\pm} &= \frac{Q_f e^2}{256 \pi^4} \sum_i  \Big[ f(z_\pm^i ) - g( z_\pm^i )  \Big]  \bar \lambda_i \tilde c_{f,i} \,,
\label{eq:haaOp_Hpm}\\
\left(\delta_f \right)^{h_i Z \gamma}_{H^\pm} &= \frac{g_{Z\bar f f}^V g_{ZH^+ H^- } }{256 \pi^4 } \Big( \frac{v}{M_{H^\pm} } \Big)^2 \sum_i  \left[
\tilde f\left(z_\pm^i\,, \frac{M_{H^\pm}^2}{m_Z^2}\right) -
\tilde g\left(z_\pm^i\,, \frac{M_{H^\pm}^2}{m_Z^2}\right) \right] \bar\lambda_{i} \tilde c_{f,i}
\label{eq:haZOp_Hpm}
\end{align}
where $z_\pm^i = M_{H^\pm}^2/ M_{h_i}^2$, $g_{ZH^+ H^- }= e(1 - \tan\theta_W^2)/(2 \tan\theta_W)$, and $\bar\lambda_i$
are the effective trilinear scalar couplings of the neutral and charged scalars, which enter the $h_i \gamma\gamma$ coupling through the $H^\pm$ loop.
Finally, the contributions from the $H^\pm W^\mp \gamma$ operators read~\cite{Abe:2013qla}
\begin{equation}
\label{eq:HpmWaOp_H}
(\delta_f)_h^{H^\pm W^\mp \gamma}= \frac{s_f}{512\pi^4} \sum_i \left[
\frac{e^2}{ 2 s_W^2} \mathcal{I}_4(M_{h_i}^2\,, M_{H^\pm}^2) \, a_i \tilde c_{f,i} -
\mathcal{I}_5 (M_{h_i}^2\,, M_{H^\pm}^2) \, \bar \lambda_i  \tilde c_{f,i} \right]\,,
\end{equation}
where $s_f=+1$ ($s_f=-1$) for the down-type quarks and charged leptons (the up-type quarks), and the two-loop integral functions are
\begin{equation}
\mathcal{I}_{4\,,5} (M_1^2\,,M_2^2) \equiv
\frac{m_W^2}{ M_{H^\pm}^2 - m_W^2 } [ I_{4\,,5}(m_W, \, M_1) - I_{4\,,5} (M_2,\,M_1)  ]\,
\end{equation}
with
\begin{align}
I_4(M_1\,, M_2)&\equiv \int_0^1 dz\, (1-z)^2 \left( z-4 + z \frac{ M_{H^\pm}^2 - M_2^2}{m_W^2}  \right)\nonumber\\
&\quad\times \frac{M_1^2}{ m_W^2 (1-z) + M_2^2 z - M_1^2 z (1-z)}
\ln \left( \frac{ m_W^2 (1-z) + M_2^2 z }{M_1^2 z (1-z) } \right),  \nonumber\\
I_5(M_1\,, M_2) &\equiv
\int_0^1 dz\, \frac{M_1^2 z(1-z)^2 }{ m_W^2 (1-z) + M_2^2 z - M_1^2 z(1-z)}\nonumber\\
&\quad\times \ln \left( \frac{ m_W^2 (1-z) + M_2^2 z }{ M_1^2 z (1-z) }\right) \,.
\end{align}
The relevant trilinear scalar couplings and the couplings among neutral Higgs bosons, charged Higgs bosons and
$W$ bosons are listed in Appendix~\ref{app:higgs_self}.

\begin{figure}[tbp]
  \begin{center}
  \includegraphics[height=0.24\textheight]{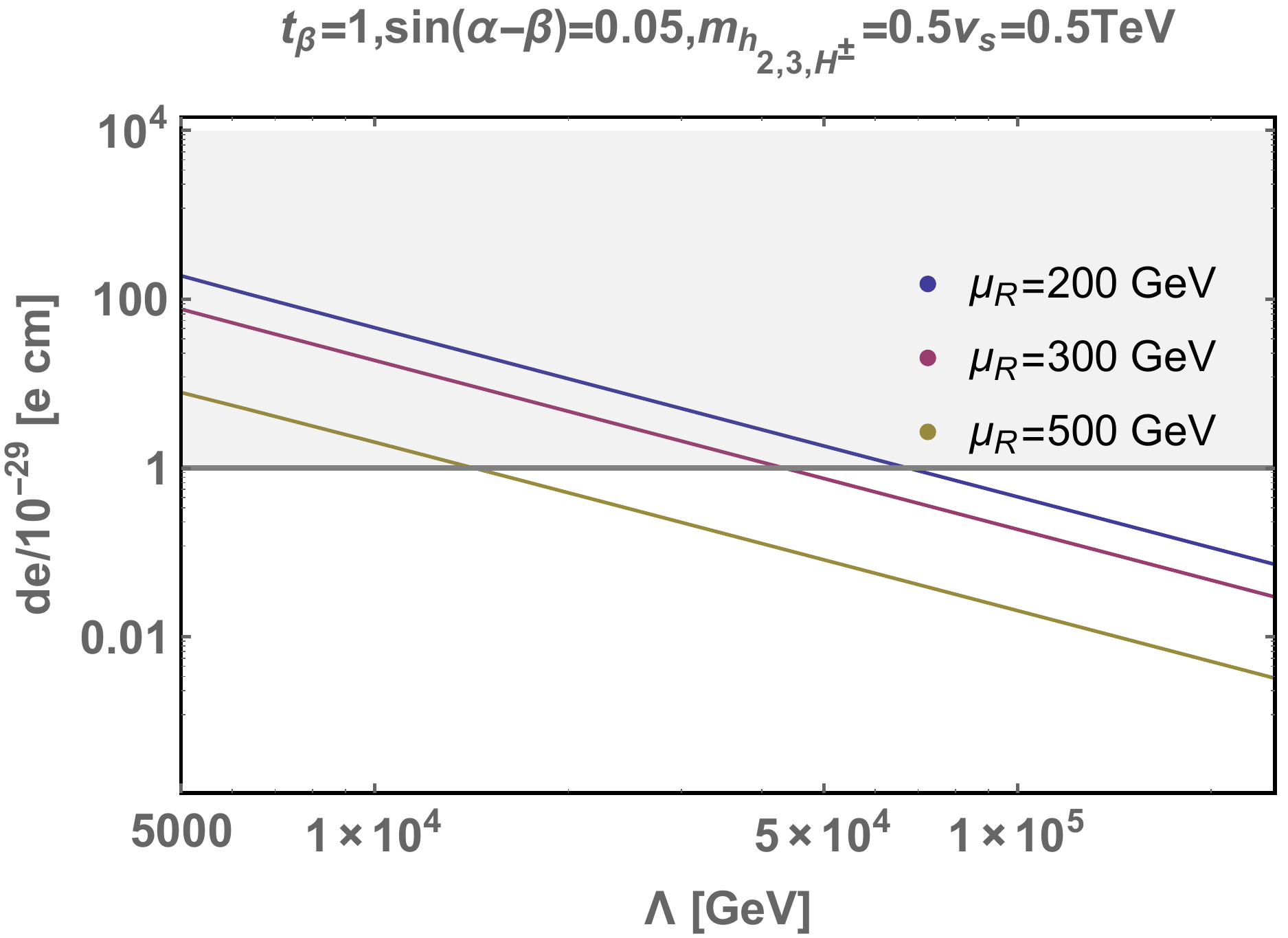}  \,\,
  \includegraphics[height=0.24\textheight]{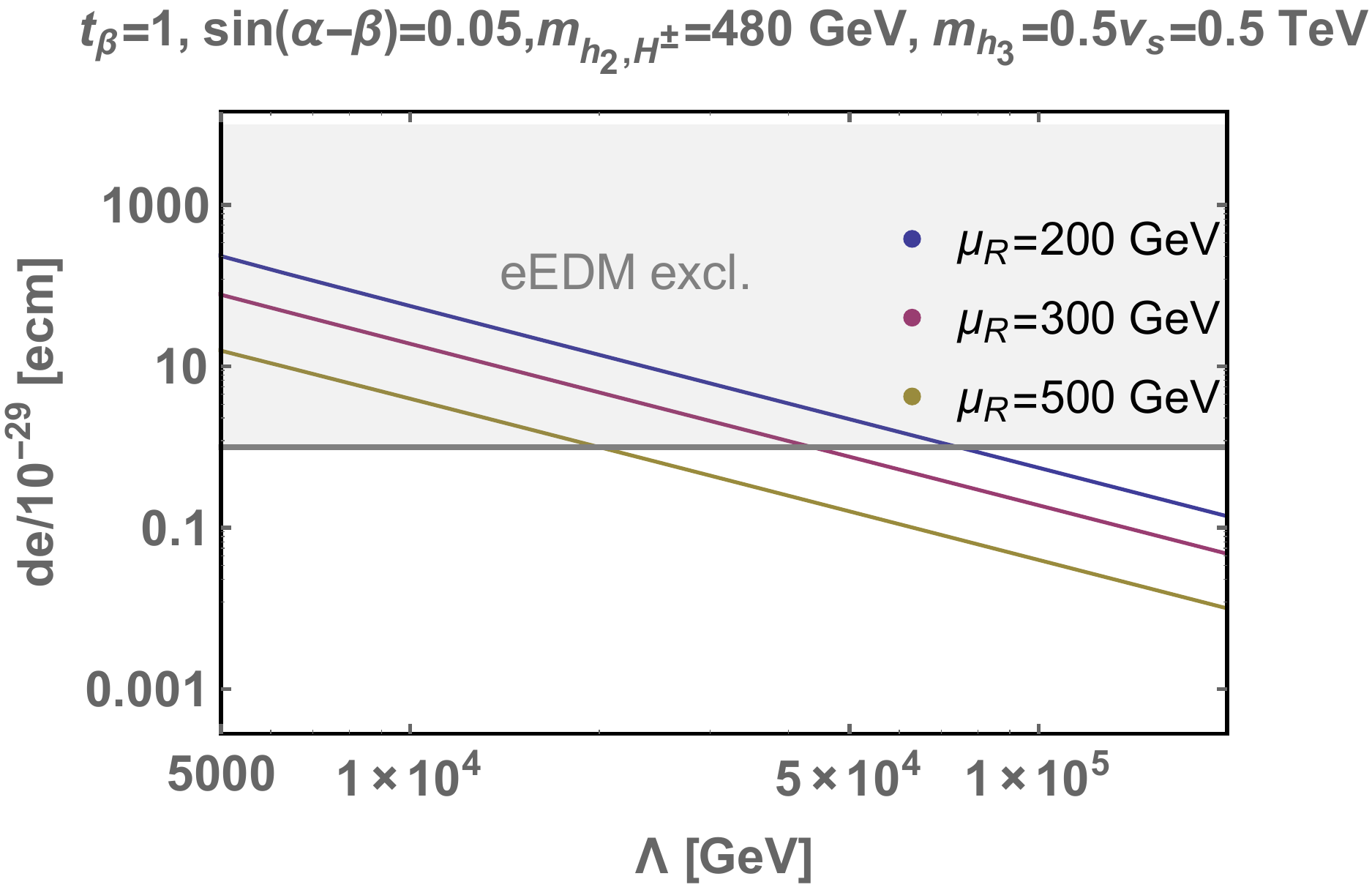}
  \end{center}
  \caption{\label{fig:de}
    The predicted value of eEDM as a function of the
    cutoff scale, for $\mu_R=200$, $300$, $500$~GeV in each panel,
    respectively. For CP-even like Higgs scalars and charged Higgs, we
    have taken $m_{h_2}=m_{h_3}=m_{H^+}=500$~GeV on the left;
    $m_{h_2}=m_{H^+}=480$~GeV and $m_{h_3}=500$~GeV on
    the right panel. We have taken $\tan\beta=1$,
    $\sin(\alpha-\beta)=0.05$, and $v_s=1$~TeV for both panels.
    The gray regions are excluded by the limit on the eEDM from ACMEII.}
\end{figure}

The eEDM contributions coming from Barr-Zee diagrams involve the neutral scalars, $h_{1,2,3,4}$, and the charged Higgs $H^\pm$.
To demonstrate the relation between the eEDM prediction and the cutoff
scale, we consider the alignment scenario with
$\sin(\alpha-\beta)=0.05$, $\tan\beta=1$, and $m_{h_2}=m_{H^+}$,
taking into account the electroweak precision bounds as studied in Refs.~\cite{Bian:2017rpg,Bian:2017xzg}.
In Fig.~\ref{fig:de}, we show the predicted value of  electron EDM as
a function of the cutoff scale, for degenerate extra scalar masses
with $m_{h_2}=m_{h_3}=m_{H^+}=500\,{\rm GeV}$ on the left, and
non-degenerate extra scalar masses with $m_{h_2}=m_{H^+}=480\,{\rm
  GeV}$ and $m_{h_3}=500\,{\rm GeV}$ on the right panel. The shaded
regions have been excluded by the bound on the electron EDM
from ACMEII\@. Here, we find that the larger  $\mu_R$, the smaller the
eEDM value for a fixed cutoff scale, which is consistent with the
CP-violating parameters $\varepsilon_i$ shown in
Fig.~\ref{fig:eps}.
The mass-degenerate case is confronted with a slightly severer bound from ACMEII\@.

\begin{figure}[tbp]
  \centering
    \includegraphics[width=0.45\textwidth]{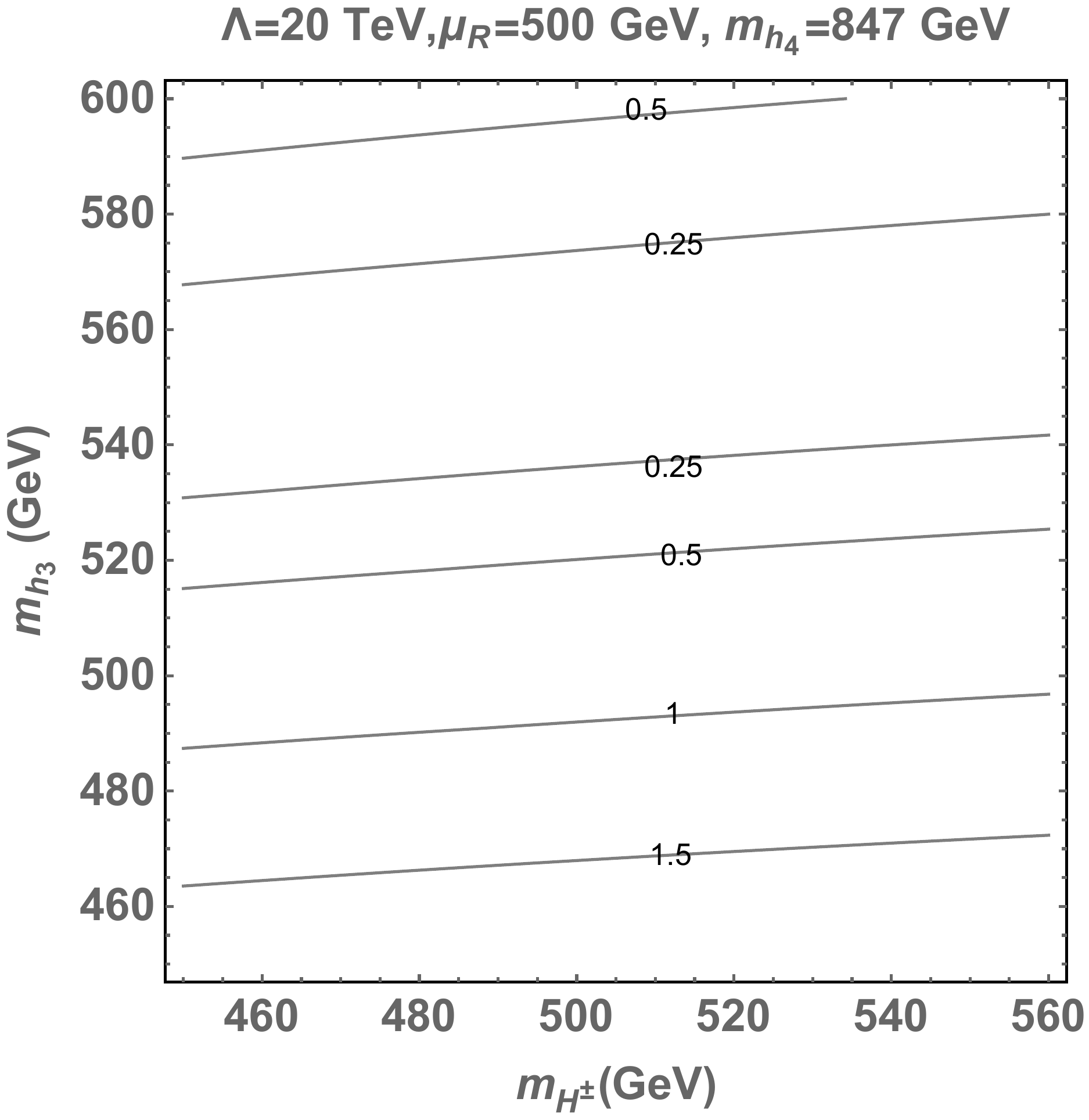} \,\,
    \includegraphics[width=0.45\textwidth]{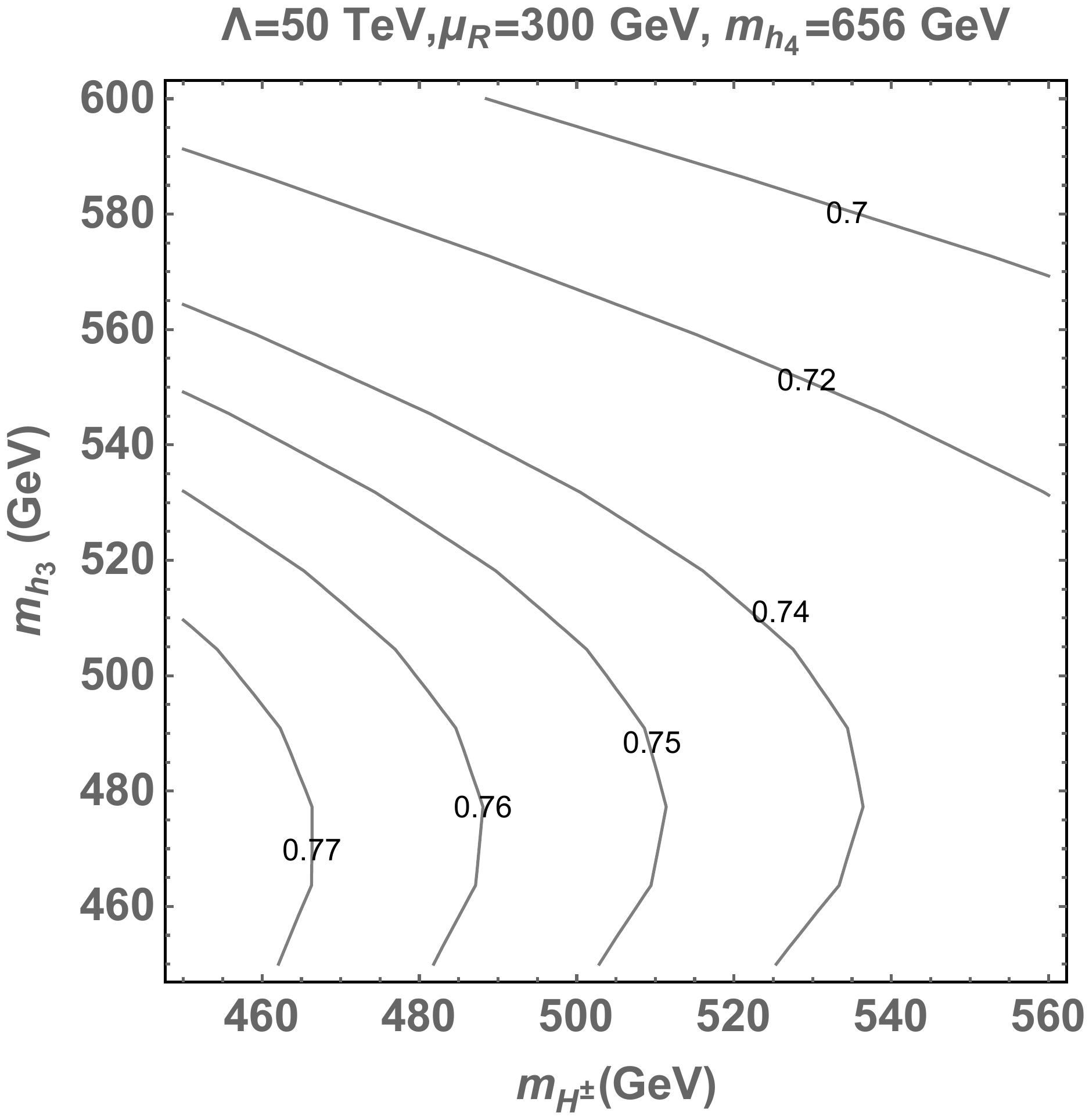}
  \caption{Contours of electron EDM in the parameter space for $m_{h_3}$ vs $m_{H^\pm}$, in units of $10^{-29}\,{\rm e\, cm}$.  }
  \label{fig:contours}
\end{figure}

To see the dependence of the eEDM predictions on heavy Higgs masses,
we show the contours of electron EDM (in units of $10^{-29}\,{\rm e\,
  cm}$) in the parameter space for singlet-like scalar ($m_{h_3}$) and
charged Higgs masses ($m_{H^\pm}$) in Fig.~\ref{fig:contours}.
In the case of $\mu_R=500$~GeV with cutoff scale $\Lambda=20$~TeV, the
lowest magnitude of the eEDM is obtained around $m_{h_3}\simeq
550$~GeV, which can be probed by the future eEDM search in ACMEIII,
as shown in the left panel. We note that the dominate contributions coming
from the $H^\pm$ loop and $H^\pm W^\mp \gamma$ cancel to some extent.
It is different from the situation studied in
Refs.~\cite{Shu:2013uua, Bian:2014zka, Bian:2016zba, Bian:2017jpt}, where the cancellation mostly occur due to top and $W$ loops. In the case of $\mu_R=300$~GeV, the current ACMEII bound becomes severer: it excludes the cutoff scale around $\Lambda\leq 50$~TeV. Therefore, we present the prediction of eEDM by choosing a larger value of the cutoff scale, $\Lambda=50$~TeV, in the right panel of Fig.~\ref{fig:contours}, so most of the parameter space for heavy Higgs masses is within the sensitivity of the ACMEIII\@.
In summary, the eEDM constraint sets the lower bound on the cutoff scale to be $\Lambda=20$--50~TeV, depending on whether the heavy Higgs masses are degenerate or not.

Before closing this section, we remark that in the parameter space of our interest, consistent with the electroweak precision data and the EDM constraints, we need relatively large masses for heavy Higgs bosons, so the contributions of charged Higgs to the lepton $g-2$ are negligible. Therefore, we do not pursue the explanation of the deviation in the muon $g-2$~\cite{amu,pdg,g2update} in our model.

\subsection{Collider searches for CP violation}

In this subsection, we discuss the independent test of the CP violation from the production of Higgs bosons at the LHC, although the bound from eEDM is already very stringent on the CP-violating mixing between neutral scalars as shown in the previous subsection. The effects of the CP violation arise in the modified Higgs couplings, parameterized by $\varepsilon_i$, in comparison with the CP-conserving case. The CP-conserving limit can be attained when $\varepsilon_i \to 0$.

In our study, we have made two assumptions in the Higgs sector. One is
that the mixing with the singlet field is negligible, so the singlet-like
Higgs boson mostly decouples in the collider phenomenology. We take
$h_3$ to be the singlet-like Higgs boson, while $h_1$ and $h_2$ are
mostly doublet-like. If the singlet-like Higgs boson is lighter than
the others, we can simply relabel the subscript.
The other assumption that we have taken is the alignment limit, where
$\sin(\alpha - \beta) \to 0$, so the non-SM-like Higgs boson $h_2$ does
not couple to the pairs of the electroweak bosons. Still, the couplings
of $h_4$ to $W^+ W^-$ and $ZZ$ do not vanish, but are proportional to
$\varepsilon_1$, in the alignment limit:
\begin{equation}
  g_{h_4 W^+ W^-} = - \frac{2 m_W^2}{v} \varepsilon_1 ,
  \quad
  g_{h_4 Z Z} = - \frac{2 m_Z^2}{v} \varepsilon_1 .
\end{equation}
However, for $|\varepsilon_1| \lesssim \mathcal{O}(10^{-3})$ and
$\Lambda > 10$~TeV, the
decays of $h_4$ to the pairs of electroweak bosons would also be
suppressed. The dominant decay modes of $h_4$ are
$h_4 \to t \bar t$ and $b \bar b$ via the $\widetilde\lambda_{t,
  b}^{h_4}$ couplings, which are independent of the $\varepsilon_i$
parameters.
Another interesting decay mode of $h_4$ is
\begin{equation}
  h_4 \to W^+ H^-
\end{equation}
if $m_{h_4} > m_W + m_{H^-}$. In the alignment limit, the coupling is
given by
\begin{equation}
  i g_{h_4 W^+ H^-}^\mu = - \frac{g}{2} \Big( N_A - i \varepsilon_2
  \Big) (p_{h_4} - p_{H^-})^\mu .
  \label{eq:coup_h4wh}
\end{equation}
The decay mode has been studied in Refs.~\cite{Haisch:2018djm,
  Kling:2018xud}, in the context of the CP-conserving 2HDM\@.
It will become more important than the $h_4 \to t \bar t$ process as
$h_4$ is heavier since the decay width is proportional to $m_{h_4}^3$,
\begin{equation}
  \Gamma (h_4 \to W^+ H^-)
  = \frac{g^2 (N_A^2 + \varepsilon_2^2)}{64 \pi m_W^2} m_{h_4}^3
  \lambda^{3/2} \left( 1, \, m_{H^-}^2 / m_{h_4}^2, \, m_W^2 /
    m_{h_4}^2 \right) ,
  \label{eq:gamma_h4wh}
\end{equation}
where $\lambda(x,\, y,\, z) = (x - y - z)^2 - 4 y z$.
Note that $\varepsilon_2$ vanishes in the alignment limit and
$\tan\beta = 1$ as can be seen in Eq.~(\ref{ep2}).
Therefore, the effect of the CP violation in the decay mode is only
relevant when we depart from the alignment limit.
Note that the same final state may appear from the decay of $h_2$.
The coupling has the similar form as in~(\ref{eq:coup_h4wh}).
However, we should fix the charged Higgs mass as $m_{H^+} = m_{h_2}$
or $m_{H^+} = m_{h_4}$ to be consistent with the constraints from the
electroweak precision~\cite{Haber:1992py, Pomarol:1993mu, Gerard:2007kn,
  Grzadkowski:2010dj, Haber:2010bw}.
By taking either choice, only one of the decay modes will be
kinematically allowed.
Due to the irreducible backgrounds of the $t \bar t$ process in the
SM, the sensitivity of the final state with $W^+ H^-$ at the LHC
turned out to be low~\cite{Haisch:2018djm}. Still, we expect that it
will be possible to probe the decay mode at the High-Luminosity LHC
and future collider experiments.

The other decay mode for the heavy Higgs bosons studied in
Refs.~\cite{Haisch:2018djm, Kling:2018xud, Kling:2020hmi} is $h_i
\to Z h_j$ for $m_{h_i} > m_Z + m_{h_j}$. The decay mode has already
been searched by the ATLAS~\cite{Aaboud:2017cxo, Aaboud:2018eoy} and
CMS~\cite{Sirunyan:2019xls, Sirunyan:2019xjg} collaborations using the final
state of $\ell^+ \ell^- + b \bar b$.
The interpretation of experimental results for the CP-conserving 2HDM
has been shown in Ref.~\cite{Kling:2020hmi}.
In the alignment limit, the coupling for the $h_4 \to Z h_1$ process
is vanishing at leading order, while the coupling at the
next-to-leading order is proportional to $\varepsilon_i^2$.
The effects of the $\varepsilon_i$ parameters for $h_4 \to Z h_2$ also
arise by the terms of the order of $\varepsilon_i^2$.
Therefore, we find that the only relevant decay mode
with the final state of $Z h_j$ for the scenario with the pure singlet
and the alignment limit is
\begin{equation}
  h_2 \to Z h_1 .
\end{equation}
The coupling in the alignment limit is
\begin{equation}
  i g_{h_1 h_2 Z}^\mu = - \frac{i N_A m_Z}{v} \varepsilon_1
  (p_{h_2} - p_{h_1})^\mu ,
\end{equation}
and the decay width is proportional to $m_{h_2}^3$, similarly as
in~(\ref{eq:gamma_h4wh}).
Therefore, searching for $h_2 \to Z h_1$ serves a direct probe of the
$\varepsilon_1$ parameter. The main background to this decay mode at
the LHC is the $Z$-boson associated Higgs production, $p p \to Z^\ast
\to Z h_1$, and the di-leptonic $t \bar t$ process in the SM\@.
We leave the detailed studies on the reach of the $\varepsilon_1$
parameter using the $h_2 \to Z h_1$ process at the LHC and
future collider experiments, as our future publication.
We stress that the other parameters, $\varepsilon_2$ and
$\varepsilon_3$, would become more relevant if we depart from the pure
singlet scenario and the alignment limit.

\section{\label{sec:UV}The UV origins of CP violation}

We discuss the origin of CP-violating higher-dimensional operators in the effective potential.
In particular, for generating the dimension-6 operator ${(S H^\dagger_1 H_2)}^2$, which captures a physical CP violation, we introduce the
NMSSM with $U(1)'$ and two other models with new doublet and singlet scalars or fermions.

\subsection{\boldmath Model A: The NMSSM with $U(1)'$ symmetry}

We consider the NMSSM with $U(1)'$ symmetry under which the singlet chiral superfield $S$ is charged (Model A).
The relevant interactions for the CP violation are given by
\begin{equation}
{\cal L}_{\rm Model\,A} =-m^2_{{\tilde t}_L}|{\tilde t}_L|^2-m^2_{{\tilde t}_R} |{\tilde t}_R|^2-y_t A_t (H^0_1)^* {\tilde t}_L {\tilde t}^*_R-y_s A_s S H_1^\dagger H_2  +{\rm h.c.}
\end{equation}
where $y_s$ is the Yukawa coupling between the singlet scalar and the
Higgsinos in the superpotential, $W=y_s S H_u H_d$, with $H_u={\tilde
  H}_1$ and $H_d=H_2$ in the basis of chiral superfields. $A_s$ and
$A_t$ are the trilinear soft mass terms.
We note that $y_s=-\mu/A_s$ in our model.
Then, from the one-loop diagram with top squarks, we get the desired
dimension-6 operator, ${(S H^\dagger_1 H_2)}^2$, with the following coefficient~\cite{UMSSM},
\begin{equation}
\frac{c_1}{\Lambda^2} = \frac{3y^4_t  y^2_s A^2_t}{32\pi^2(m^2_{{\tilde t}_1}-m^2_{{\tilde t}_2})^2}\, {\cal G}(m^2_{{\tilde t}_1},m^2_{{\tilde t}_2}) \label{matching}
\end{equation}
where $m^2_{{\tilde t}_{1,2}}$ are the squared masses for top squarks, and the loop function $\cal G$ is given by
\begin{equation}
{\cal G}(m^2_{{\tilde t}_1},m^2_{{\tilde t}_2})=2- \frac{m^2_{{\tilde t}_1}+m^2_{{\tilde t}_2}}{m^2_{{\tilde t}_1}-m^2_{{\tilde t}_2}} \,\ln\left(\frac{m^2_{{\tilde t}_1}}{m^2_{{\tilde t}_2}} \right).
\end{equation}
For $m^2_{{\tilde t}_2}\gg m^2_{{\tilde t}_1}\gg |A_t| m_t $, we can approximate Eq.~(\ref{matching}) as
\begin{equation}
\frac{c_1}{\Lambda^2}  \approx  \frac{3y^4_t  y^2_s A^2_t}{32\pi^2 m^4_{{\tilde t}_2}}\,\left[ 2+\ln\bigg(\frac{m^2_{{\tilde t}_1}}{m^2_{{\tilde t}_2}} \bigg) \right].
\end{equation}
In this case, the nontrivial CP phase in the dimension-6 operator is originated from the CP phase in $A_t$.

\begin{figure}[tbp]
  \begin{center}
    \includegraphics[width=0.45\textwidth]{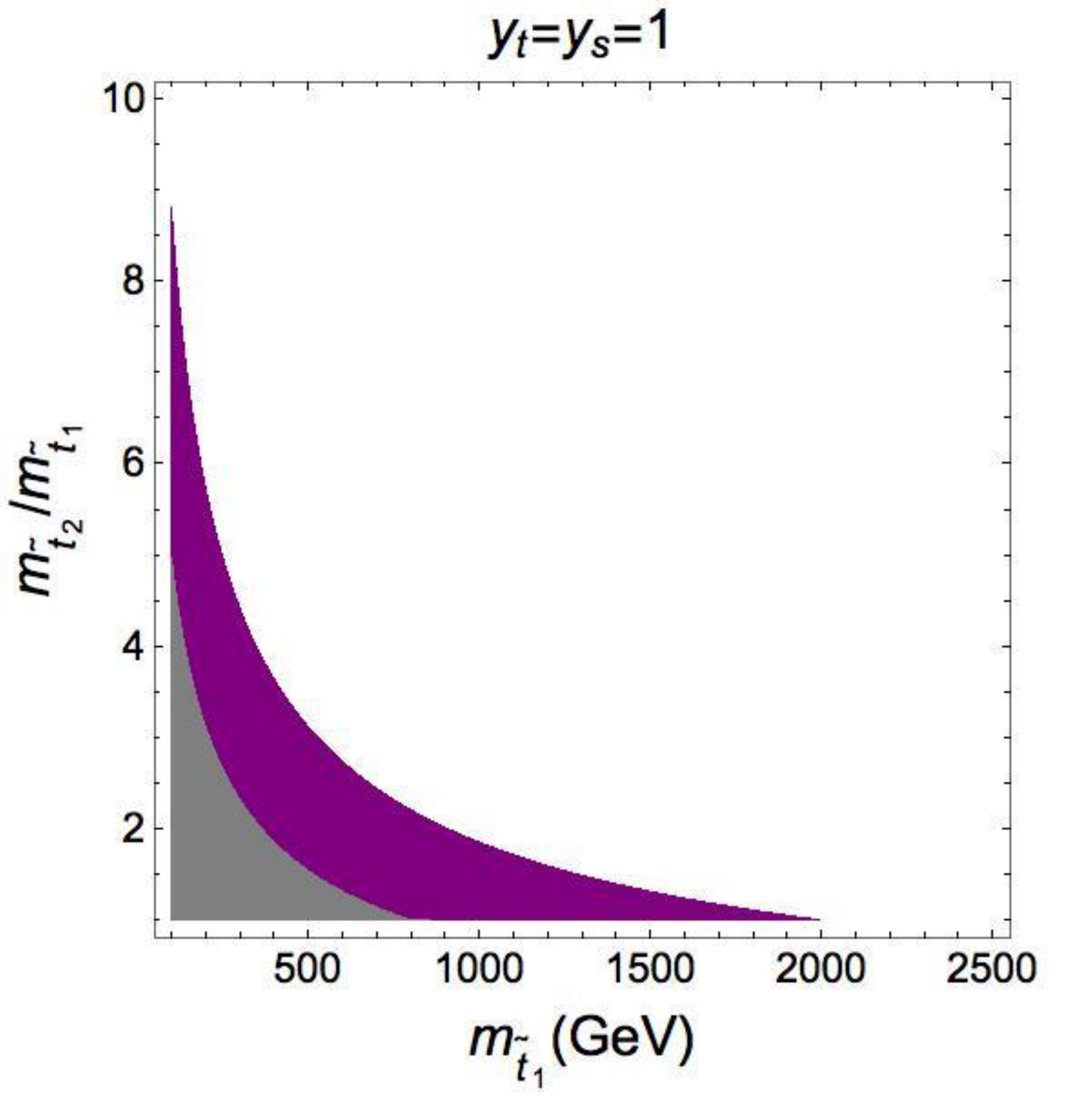}
  \end{center}
    \caption{\label{fig:stops}
      The parameter space for stop masses, $m_{{\tilde t}_1}$ and
    $m_{{\tilde t}_2}/m_{{\tilde t}_1}$, with the eEDM constraints. We
    have set $A_t=m_{{\tilde t}_1}$, $y_t=y_s=1$ and ${\rm Arg}(y_t^4
    y^2_s A^2_t)=\pi/2$. The regions colored in gray and magenta are
    ruled out by the eEDM bounds, for $\Lambda/\sqrt{|c_I|}<20$ and
    $50$~TeV, respectively.}
\end{figure}

In view of the eEDM constraints discussed for the cutoff scale in the previous section, we can impose $\Lambda=20$--$50$~TeV depending on the masses of heavy Higgs bosons, which can be translated to the bounds on the stop masses and mixing parameter. In Fig.~\ref{fig:stops}, we show the parameter space for stop masses, $m_{{\tilde t}_1}$ and $m_{{\tilde t}_2}/m_{{\tilde t}_1}$, which is ruled out by the eEDM bounds with the cutoff scale greater than $20$, $50$~TeV colored in gray and magenta, respectively. Here we have taken $y_t=y_s= 1$, $A_t=m_{{\tilde t}_1}$ and  ${\rm Arg}(y_t^4 y^2_s A^2_t)=\pi/2$.
Therefore, for mass-degenerate heavy Higgs bosons, for which
the cutoff scale is constrained to be greater than $50\,{\rm TeV}$,
the lighter stop mass should be larger than up to $2\,{\rm TeV}$
(colored in magenta in Fig.~\ref{fig:stops}), depending on the mass of the heavier stop mass. As a result, we may probe the stop masses with the eEDM measurement beyond the reach of the LHC\@.
On the other hand, for non-degenerate masses of heavy Higgs bosons,
for which the cutoff scale is constrained to be as low as 20~TeV, the
lighter stop mass up to $800\,{\rm GeV}$ (colored in gray in
Fig.~\ref{fig:stops}) is ruled out by the eEDM bound.
In the CP-violating NMSSM with CP violation from both tree and
loop-levels but with no $U(1)'$, we refer to Ref.~\cite{Bian:2017wfv}
for a complete study on baryon asymmetry of the Universe and the EDM\@.

\subsection{Model B: models with doublet and singlet scalars}

Another example worth considering is the scalar dark matter as the origin of the CP violation (Model B).
For this, we introduce a SM doublet $\phi_D$ with hypercharge $Y=+\frac{1}{2}$ and a SM singlet $\phi_S$, which are neutral under the $U(1)'$ and the SM color.
We also impose a global symmetry, ${U(1)}_R$, under which $S$ carries
charge $+2$, $\phi_D, \phi_S$ carry charge $+1$ whereas two Higgs
doublets and the SM fermions are neutral, as shown in
Table~\ref{tab:model_B}. The ${U(1)}_R$ symmetry corresponds to the one
in the supersymmetric models as in Model A where the $A$-term softly
breaks the ${U(1)}_R$ symmetry.

In this setup, we introduce the couplings between the extra scalars
and $S, H_{1,2}$, as in the following Lagrangian,
\begin{equation}
{\cal L}_{\rm Model\,B} =-m^2_{D} |\phi_D|^2-m^2_S|\phi_S|^2
-\lambda_D S H_2\, \phi^\dagger_D \phi^*_S - A_D H^\dagger_1 \phi_D
\phi_S +{\rm h.c.} ,
\end{equation}
where $A_D$ is the spurion parameter carrying charge $-2$ under the ${U(1)}_R$.
The ${U(1)}_R$ symmetry is softly broken to $Z_2$ due to the $A_D$ term
as well as the $\mu$ term. Then, since $\phi_D$ and $\phi_S$  are
$Z_2$ odd, the lighter neutral complex scalar among them can be a dark matter candidate.
Then, similarly as the previous example, from the loops with new scalars, we can obtain the dimension-6 operator $(S H^\dagger_1 H_2)^2$, with the coefficient, $\frac{c_1}{\Lambda^2}\sim \frac{\lambda^2_D  A^2_D}{16\pi^2 m^4_D}$ for $m_D\gg m_S\gg A_D v_1$.
In this case, the nontrivial CP phase in the dimension-6 operator stems from the CP phase
of the coupling to the dark scalars $A_D$.

\begin{table}[hbt!]
  \begin{center}
    \begin{tabular}{c|cccccc}
      \hline\hline
      &&&&&&\\[-2mm]
      & $S$  & $H_1$ & $H_2$ & $\phi_S$ & $\phi_D$ & $A_D$\\[2mm]
      \hline
      &&&&&&\\[-2mm]
      $Q'$ &  $\frac{1}{3}x$ & $0$ & $-\frac{1}{3}x$  & $0$ & $0$
                     & $0$\\[2mm]
      \hline
      &&&&&&\\[-2mm]
      $U(1)_R$ &  $+2$ & $0$ & $0$  & $+1$ & $+1$
                     & $-2$\\[2mm]
      \hline\hline
    \end{tabular}
  \end{center}
  \caption{\label{tab:model_B}$U(1)'$ and $U(1)_R$ charges of scalars for Model B.}
\end{table}

There are similar bounds on the masses for doublet and singlet scalars, $m_D$ and $m_S$, similarly as in the NMSSM with $U(1)'$, if we identify $\lambda_D=y^2_t y_s$ and $A_D\sim A_t\sim m_S$.
The difference from the NMSSM with $U(1)'$ is that new particles running in the loops contain charge-neutral scalars, the lighter of which can be a dark matter candidate, that is, $\phi_S$, unlike the stops. Therefore, it would be interesting to pursue the details on the interplay between the eEDM bound and the CP violation in the dark sector.

 \subsection{Model C: models with doublet and singlet fermions}

We also consider the possibility of fermion dark matter for the CP violation (Model C).
We introduce a vector-like doublet fermion, composed of $\psi,
{\widetilde\psi}$, with hypercharge $Y=-\frac{1}{2}, +\frac{1}{2}$,
and a Weyl singlet fermion $\psi'$, that are neutral under the
$U(1)'$, and a vector-like singlet fermion, composed of $\chi,
{\widetilde\chi}$, which carry charges $-\frac{1}{3}x, +\frac{1}{3}x$
under the $U(1)'$. We also assign the charges for scalars and
two-component spinors under the global ${U(1)}_R$ symmetry as in
Table~\ref{tab:model_C}.
Thus, as in the previous models, the ${U(1)}_R$ is softly broken to $Z_2$ by the $\mu$ term and the Dirac mass term for $\chi, {\widetilde\chi}$. Then, the lightest neutral fermion among the extra neutral fermions, which are $Z_2$ odd, can be a dark matter candidate.

\begin{table}[hbt!]
  \begin{center}
 \begin{tabular}{c|cccccccc}
      \hline\hline
      &&&&&&&&\\[-2mm]
      & $S$  & $H_1$ & $H_2$ & $\psi'$ & $\psi$ & $\widetilde\psi$ & $\chi$ & $\widetilde\chi$ \\[2mm]
      \hline
      &&&&&&&&\\[-2mm]
      $Q'$ &  $\frac{1}{3}x$ & $0$ & $-\frac{1}{3}x$  & $0$ & $0$ & $0$
                     & $-\frac{1}{3}x$ & $+\frac{1}{3}x$ \\[2mm]
      \hline
      &&&&&&&&\\[-2mm]
 $U(1)_R$ &  $+2$ & $0$ & $0$  & $-1$ & $+1$ & $-1$
                     & $-1$ & $-1$ \\[2mm]
      \hline\hline
    \end{tabular}
  \end{center}
    \caption{\label{tab:model_C}$U(1)'$ and $U(1)_R$ charges of scalars and extra fermions for Model C.}
\end{table}

Then, in the two-component spinor notations, the Lagrangian for the
extra fermions is given by
\begin{equation}
{\cal L}_{\rm Model\,C} =-m_\psi \psi {\widetilde\psi}-m_\chi
\chi^\dagger {\widetilde\chi}^\dagger- \lambda_S \, S \psi'\chi - y_1
{\tilde H}_1 \psi^\dagger\psi^{\prime\dagger} - y_2 H_2 {\widetilde
  \chi} \,\psi +{\rm h.c.} ,
\end{equation}
where $m_\chi$ is the spurion mass parameter carrying charge $-2$ under the ${U(1)}_R$.
Due to the loops with extra fermions, the dimension-6 operator  ${(S H^\dagger_1 H_2)}^2$ is generated with the coefficient, $\frac{c_1}{\Lambda^2}\sim \frac{\lambda^2_S y^2_1 y^2_2 m^2_\chi}{16\pi^2 m^4_\psi}$.
In this case, the nontrivial CP phase in the dimension-6 operator is from the CP phases
of the Yukawa couplings to the extra fermions and/or the Dirac mass term, and it can be sufficiently suppressed for $m_\psi\gg m_\chi$ with Yukawa couplings, $\lambda_S, y_{1,2}$, being of order one.

As compared to the case in the NMSSM with $U(1)'$, the role of dimensionless couplings is played by  $\lambda_S y_1 y_2=y^2_t y_s$ and the dimensionful parameter is translated to $m_\chi\sim A_t$.
The smallness of the CP-violating dimension-6 operator is
attributed to a small ${U(1)}_R$ breaking mass term for $\chi$ and
${\widetilde\chi}$.  While the $\psi, {\widetilde\psi}$ and
$\psi',\chi$ pairs have large Dirac masses, ${\widetilde\chi}$, having
a Majorana fermion with a small mass, is a candidate for dark
matter. We postpone the detail analysis of the model in a future
publication.

\section{Conclusions}

We have presented a parametrization of the CP violation in the effective theory for the 2HDM with a local $U(1)'$ and showed how the higher-dimensional operators in the scalar potential  violate the CP symmetry at the observable level. The tadpole condition from the minimization of the scalar potential renders the Higgs mixing mass parameter carry a nonzero CP phase by the interplay with the higher-dimensional operators. We calculated the EDM of the electron arising at two loops due to the mixings among CP-even and CP-odd scalars in our model and identified the cutoff scale from the bound on eEDM to be $20$--$50$~TeV, depending on the mass spectrum of heavy Higgs bosons.
The results are applicable to general 2HDMs where the Higgs mixing
mass term is generated by the $U(1)'$ breakdown, provided that there
is no significant violation of flavor in the Yukawa couplings for
charged leptons.

We have shown how the inputs from the collider searches for heavy Higgs bosons with CP violation can be used to make an independent test of the CP-violating parameters in the models. In particular, in the alignment limit favored by the 125~GeV Higgs data, the pseudoscalar-like scalar can also decay into $WW$ or $ZZ$, and it has  the $W^+H^-$ decay mode modified due to the CP-violating mixing parameters, which might be testable at the High-Luminosity LHC\@. Furthermore, the $h_2\rightarrow Z h_1$  process with $\ell^+ \ell^-+b{\bar b}$ final states at the LHC and future collider experiments can serve a direct probe of the CP-violating parameter against the backgrounds coming from the $Z$-boson associated production of $h_1$ or the di-leptons from $t{\bar t}$.

We have also discussed the microscopic origins for generating the higher-dimensional operators in the scalar potential, in the context of both supersymmetric or non-supersymmetric models. In the case of the NMSSM with $U(1)'$, the mass parameters of stops running in loops can generate the CP-violating dimension-6 operator in the scalar potential, thus they can be constrained indirectly by the eEDM according to our general results.   Depending on whether the heavy Higgs bosons have split masses or not, the eEDM bound can constrain the lighter stop mass to be heavier than up to $800\,{\rm GeV}$--$2\,{\rm TeV}$, thus being complementary to the direct searches for stops at the High-Luminosity LHC and future colliders. In models with new neutral scalars or fermions running in loops, the lighter neutral particle is a good candidate for dark matter with CP-violating couplings, so there can be a variety of  ways of probing the CP violation by dark matter experiments as well as more precise measurements of eEDM such as ACMEIII\@.

Finally we remark that in CP-violating 2HDMs, there is a tension between the strong signal for gravitational waves and the electroweak baryogenesis~\cite{Zhou:2020xqi,Dorsch:2016nrg}. We leave the possibility of addressing the baryon asymmetry of the Universe in microscopic models with $U(1)'$ to a future study.

\section*{Acknowledgments}

The work of LB is supported by the National Natural Science Foundation of China under grant No.11605016 and No.11947406.
The work of HML is supported in part by Basic Science Research Program
through the National Research Foundation of Korea (NRF) funded by the
Ministry of Education, Science and Technology (NRF-2019R1A2C2003738 and NRF-2018R1A4A1025334).
CBP is supported by IBS under the project code, IBS-R018-D1.

\appendices%

\section{The minimization and tadpole conditions\label{app:higgs_sector}}

Although we have not considered the $\Phi_a$ fields, responsible
for generating neutrino masses~\cite{Bian:2017rpg,Bian:2017xzg}, in our
study, we here take them into account for the sake of completeness.
The full scalar potential is now composed of $V = V_1 + \Delta V_1 +
V_2 + \Delta V_2$,
with $V_1$ and $V_2$ given in Eqs.~(\ref{s1}) and (\ref{s30}), respectively.
$\Delta V_1$ contains the renormalizable
terms containing the extra singlet scalar fields $\Phi_a$,
and $\Delta V_2$ does the additional higher-dimensional operators
due to singlet scalar fields as follows:
\begin{align}
  \Delta V_1=
  &~\sum_{a=1}^3\Big( \mu^{\prime 2}_{\Phi_a} |\Phi'_i|^2
    +\lambda'_{\Phi_a}|\Phi'_a|^4 \Big)+ \left(\rho' S^3
    \Phi^{\prime\dagger}_3 +\mu'_\Phi \Phi'_1 \Phi'_2 \Phi^{\prime\dagger}_3
    + \mathrm{h.c.} \right) \nonumber \\
  &+ 2\sum_{a=1}^3 |\Phi'_a|^2(\beta'_{a1} |H'_1|^2 +\beta'_{a2}  |H'_2|^2+
    \beta_{a3} |S'|^2)+2 \sum_{a<b}\lambda'_{ab} |\Phi'_a|^2 |\Phi'_b|^2,  \label{s2}\\
   \Delta  V_2=
    &~ \frac{c'_2}{\Lambda^2}\, S^{\prime\dagger} \Phi'_3 (H^{\prime\dagger}_1 H'_2)^2 +\frac{d'_1}{\Lambda}\, S^{\prime 3} \Phi'_1 \Phi'_2 + \frac{d'_2}{\Lambda^2}\, (\Phi'_1\Phi'_2\Phi^{\prime\dagger}_3)^2 +{\rm h.c.} +\cdots. \label{s3}
\end{align}
Here we have kept up to dimension-6 terms in the potential $\Delta V_2$ and the ellipses denote even higher-dimensional terms.
Then, in this model, there are new CP phases from $\mu'$, $\rho'$,
$\mu'_\Phi$ as well as $c'_1$, $c'_2$, $d'_1$, and $d'_2$.

For the scalar potential with phase-rotated scalar fields  and
redefined parameters, the minimization conditions yield
\begin{align}
  \mu_1^2 =
  &~ \frac{1}{\sqrt{2}} \Re (\mu )
    \frac{v_2 v_s}{v_1} - \lambda_1 v_1^2 - (\lambda_3 + \lambda_4)
    v_2^2 - \kappa_1 v_s^2 + \sum_{a=1}^3 \beta_{a 1} \omega_a^2
    \nonumber \\
  & - \frac{1}{2 \Lambda^2} \Re (c_1 ) v_2^2 v_s^2 - \frac{1}{2 \Lambda^2} \Re (c_2 ) v_2^2 v_s \omega_3 , \\
  \mu_2^2 =
  &~ \frac{1}{\sqrt{2}} \Re (\mu )
    \frac{v_1 v_s}{v_2} - \lambda_2 v_2^2 - (\lambda_3 + \lambda_4)
    v_1^2 - \kappa_2 v_s^2 + \sum_{a=1}^3 \beta_{a 2} \omega_a^2
    \nonumber\\
  & - \frac{1}{2 \Lambda^2} \Re (c_1 ) v_1^2 v_s^2 - \frac{1}{2 \Lambda^2} \Re (c_2 ) v_1^2 v_s \omega_3, \\
  m_S^2 =
  &~ \frac{1}{\sqrt{2}} \Re (\mu )
    \frac{v_1 v_2}{v_s} - \lambda_S v_s^2 - \kappa_1 v_1^2 -\kappa_2
    v_2^2  - \sum_{a=1}^3 \beta_{a 3} \omega_a^2
    - \frac{3}{2} \Re (\rho) v_s \omega_3
    \nonumber\\
  & -\frac{3}{2 \sqrt{2} \Lambda} \Re (d_1 ) v_s \omega_1 \omega_2  - \frac{1}{2 \Lambda^2} \Re (c_1 ) v_1^2 v_2^2
    - \frac{1}{4 \Lambda^2} \Re (c_2 ) \frac{v_1^2 v_2^2 \omega_3}{v_s}, \\
  \mu_{\Phi_1}^2 =
  & - \frac{1}{\sqrt{2}} \Re (\mu_\Phi )
    \frac{\omega_2 \omega_3}{\omega_1} - \lambda_{\Phi_1} \omega_1^2
    - \left( \beta_{11} v_1^2 + \beta_{12} v_2^2 + \beta_{13} v_s^2
    \right)
    - \lambda_{12} \omega_2^2 - \lambda_{13} \omega_3^2 \nonumber\\
  & - \frac{1}{2\sqrt{2} \Lambda} \Re (d_1 ) \frac{v_s^3 \omega_2}{\omega_1}
    - \frac{1}{2 \Lambda^2} \Re(d_2 )
    \omega_2^2 \omega_3^2 , \\
  \mu_{\Phi_2}^2 =
  & - \frac{1}{\sqrt{2}} \Re (\mu_\Phi)
    \frac{\omega_1 \omega_3}{\omega_2} - \lambda_{\Phi_2} \omega_2^2
    - \left( \beta_{21} v_1^2 + \beta_{22} v_2^2 + \beta_{23} v_s^2
    \right)
    - \lambda_{12} \omega_1^2 - \lambda_{23} \omega_3^2 \nonumber\\
  & - \frac{1}{2 \sqrt{2} \Lambda} \Re (d_1 ) \frac{v_s^3 \omega_1}{\omega_2}
    - \frac{1}{2 \Lambda^2} \Re (d_2 )
    \omega_1^2 \omega_3^2 , \\
  \mu_{\Phi_3}^2 =
  & - \frac{1}{2} \Re( \rho)
    \frac{v_s^3}{\omega_3}
    - \frac{1}{\sqrt{2}} \Re (\mu_\Phi)
    \frac{\omega_1 \omega_2}{\omega_3} - \lambda_{\Phi_2} \omega_3^2
    \nonumber\\
  & - \left( \beta_{31} v_1^2 + \beta_{32} v_2^2 + \beta_{33} v_s^2
    \right) - \lambda_{13} \omega_1^2 - \lambda_{23} \omega_2^2
    \nonumber\\
  & - \frac{1}{4 \Lambda^2} \Re(c_2 ) \frac{v_1^2 v_2^2 v_s}{\omega_3}
    - \frac{1}{2 \Lambda^2} \Re(d_2 )
    \omega_1^2 \omega_2^2 .
\end{align}

The tadpole parameters for the pseudoscalar fields are
given by
\begin{align}
  \frac{T_{\eta_1}}{v_2} = - \frac{T_{\eta_2}}{v_1} =
  & - \frac{1}{\sqrt{2}} \Im (\mu ) v_s + \frac{1}{2 \Lambda^2} \Im (c_1 ) v_1 v_2 v_s^2
    + \frac{1}{2 \Lambda^2} \Im (c_2 ) v_1 v_2 v_s \omega_3 , \\
  \frac{T_{\eta_S}}{v_s} =
  &~ \frac{1}{\sqrt{2}} \Im (\mu ) \frac{v_1 v_2}{v_s} - \frac{3}{2} \Im (\rho ) v_s \omega_3 - \frac{3}{2 \sqrt{2} \Lambda} \Im (d_1 ) v_s \omega_1 \omega_2 \nonumber\\
    & - \frac{1}{2 \Lambda^2} \Im (c_1 ) v_1^2 v_2^2 + \frac{1}{4 \Lambda^2} \Im (c_2
      ) \frac{v_1^2
      v_2^2 \omega_3}{v_s} , \\
  \frac{T_{\Phi_{1I}}}{\omega_2} = \frac{T_{\Phi_{2I}}}{\omega_1} =
  & - \frac{1}{\sqrt{2}} \Im(\mu_\Phi ) \omega_3  - \frac{1}{2 \sqrt{2} \Lambda} \Im(d_1 ) v_s^3
      - \frac{1}{2 \Lambda^2} \Im (d_2) \omega_1 \omega_2 \omega_3^2 , \\
  \frac{T_{\Phi_{3I}}}{\omega_3} =
  &~ \frac{1}{\sqrt{2}} \Im (\mu_\Phi ) \frac{\omega_1 \omega_2}{\omega_3}
    + \frac{1}{2} \Im (\rho )
    \frac{v_s^3}{\omega_3} \nonumber\\
  & - \frac{1}{4 \Lambda^2} \Im (c_2 ) \frac{v_1^2 v_2^2 v_s}{\omega_3}
    + \frac{1}{2 \Lambda^2} \Im (d_2 )
    \omega_1^2 \omega_2^2 .
\end{align}
By combining the above relations, we have
\begin{align}
  0
  &= \frac{v_1 v_2}{v_s^3} \frac{T_{\eta_1}}{v_2}
    + \frac{T_{\eta_S}}{v_s}
    + \frac{3 \omega_1 \omega_2}{v_s^2} \frac{T_{\Phi_{1I}}}{\omega_2}
    + \frac{3 \omega_3^2}{v_s^2} \frac{T_{\Phi_{3I}}}{\omega_3} \nonumber\\
  &= - \frac{3}{\sqrt{2} \Lambda} \Im (d_1 ) v_s \omega_1 \omega_2 ,
\end{align}
so $\Im (d_1) = 0$. Then, we
further find that
\begin{align}
  0
  &= \frac{v_1 v_2}{3 v_s^3} \frac{T_{\eta_1}}{v_2}
  + \frac{1}{3} \frac{T_{\eta_S}}{v_s} \nonumber\\
  &= - \frac{1}{2} \Im (\rho ) v_s \omega_3
  + \frac{1}{4 \Lambda^2} \Im (c_2 ) \frac{v_1^2 v_2^2 \omega_3}{v_s},
\end{align}
which results in $\Im (\rho ) = \Im (c_2 ) = 0$. The remaining
combinations are
\begin{align}
  0 &= \left . \frac{T_{\eta_1}}{v_2} \right\vert_{\Im (c_2 ) = 0} \nonumber\\
  &= - \frac{1}{\sqrt{2}} \Im (\mu) v_s + \frac{1}{2 \Lambda^2} \Im (c_1 ) v_1 v_2 v_s^2 ,  \label{tad1} \\
  0 &= \left . \frac{T_{\Phi_{1I}}}{\omega_2} \right\vert_{\Im (d_1 ) = 0} \nonumber\\
    &= - \frac{1}{\sqrt{2}} \Im(\mu_\Phi ) \omega_3
    - \frac{1}{2 \Lambda^2} \Im (d_2 ) \omega_1 \omega_2 \omega_3^2 . \label{tad2}
\end{align}
Consequently, the tadpole conditions in eqs.~(\ref{tad1}) and
(\ref{tad2}) determine the CP phases of the mass parameters, $\mu$ and
$\mu_\Phi$, in terms of the CP phases of the dimension-6 operators,
$c_1$ and $d_2$, respectively. In the text, we used the results
focusing on $c_1$ in order to see the CP-violating mixings between
CP-even and CP-odd scalars in the 2HDMs with $U(1)'$.

\section{Diagonalization of scalar mass matrices \label{app:higgs_mixing}}

In the absence of the CP violation,  the would-be Goldstone bosons,
$G_Y$ and $G'$, for the spontaneously broken $U(1)_Y\times U(1)'$,  can be identified as follows,
\begin{align}
G_Y &= \frac{2}{v} \Big(\frac{1}{2}v_1\eta_1 + \frac{1}{2} v_2\eta_2 \Big)=\cos\beta\,\eta_1+\sin\beta\, \eta_2, \label{GY} \\
G'&= \frac{3}{x v_{Z'}} \Big(\frac{1}{3}x v_s S_I  -\frac{1}{3} x v_2 \eta_2\Big)=\frac{1}{v_{Z'}} \Big(v_s S_I - v\sin\beta\,\eta_2\Big), \label{Gp}
\end{align}
with $v^2=v^2_1+v^2_2$ and $v^2_{Z'}\equiv v^2_s + v^2_2$. Here, we note that the $Z'$ gauge boson is given
by $m_{Z'}\simeq \frac{1}{3}x g_{Z'} v_{Z'}$ if the extra singlet VEVs from $\Phi_a$ are small.
Then, the heavy pseudoscalar $A^0$ can be taken to be orthogonal to the above two would-be Goldstone bosons as
\begin{equation}
A^0 =N_A \left(\sin\beta\, \eta_1 -\cos\beta\, \eta_2-\frac{v}{v_s}\,\sin\beta\cos\beta\, S_I \right) \label{A0}
\end{equation}
with $N_A$ given in (\ref{eq:NA}).
Therefore, we now make a transformation to the basis with would-be
Goldstone bosons, $G_Y$ and $G'$, and the heavy pseudoscalar $A^0$, by
\begin{equation}
  \begin{pmatrix}
    \eta_1 \\ \eta_2 \\ S_I
  \end{pmatrix} =
  \mathcal{R}_3
  \begin{pmatrix}
    A^0 \\ G_Y \\ G^\prime
  \end{pmatrix} , \label{goldstones}
\end{equation}
where $ {\cal R}_{3}$ is the $3\times 3$ rotation matrix and its
inverse reads from Eqs.~(\ref{GY}), (\ref{Gp}) and (\ref{A0}) as
follows.
\begin{equation}
{\cal R}_3=\left(\begin{array}{ccc}  \frac{N_A v_2}{v}  &  -\frac{N_A v_1}{v} &  -\frac{N_A v_1 v_2}{v v_s}   \vspace{0.2cm} \\ \frac{v_1}{v} & \frac{v_2}{v} & 0 \vspace{0.2cm}\\  0 &  - \frac{v_2}{v_{Z'}} &  \frac{v_s}{v_{Z'}} \end{array}\right)^{-1}. \label{r3}
\end{equation}
We find that $G_Y$ and $G'$ appear massless as expected, and $A^0$
mixes with the CP-even scalars due to the CP violation.
The results are used to choose the basis for the squared mass matrix
for scalars with CP violation in the text.

We note from Eq.~(\ref{goldstones}) that  the CP-odd scalars can be expressed in terms of the CP-odd scalar $A^0$ and the would-be Goldstone bosons as
\begin{align}
\eta_1 &= N_A \Big(  \sin\beta\, A^0 + \frac{N_A v^2_{Z'}}{v^2_s}\, \cos\beta\, G_Y + \frac{N_A v v_{Z'}}{v^2_s}\, \cos\beta\sin^2\beta\, G'\Big),  \label{ps1} \\
\eta_2 &= N_A \Big(  -\cos\beta\, A^0 + N_A \sin\beta\, G_Y - \frac{N_A v v_{Z'}}{v^2_s}\,\cos^2\beta\sin\beta\, G'\Big),\label{ps2}  \\
S_I &= N_A \Big(  -\frac{v}{v_s}\, \sin\beta\cos\beta\, A^0 + \frac{N_A v}{v_s}\, \sin^2\beta\, G_Y + \frac{N_A v_{Z'}}{v_s}\, G'\Big). \label{ps3}
\end{align}
Working in the basis of $(\rho_1,\rho_2,S_R, A^0)$, where $A^0$ is the CP-odd Higgs in the absence of CP violation,
the squared mass matrix is given in the following $4\times 4$ matrix
form, $M^2_0$, with
\begin{align}
(M^2_0)_{11} &=   2\lambda_1 v^2_1 +\frac{\mu_R v_2 v_s}{\sqrt{2} v_1},   \nonumber\\
(M^2_0)_{12} &=   (M^2_0)_{21}=2 v_1 v_2 (\lambda_3+\lambda_4)-\frac{\mu_R v_s}{\sqrt{2}}+\frac{c_R}{\Lambda^2}\,v_1v_2 v_s^2 , \nonumber\\
(M^2_0)_{13} &=  (M^2_0)_{31}=2\kappa_1 v_1 v_s-\frac{\mu_R v_2}{\sqrt{2}} +\frac{c_R}{\Lambda^2}\,v_1 v_2^2 v_s,   \nonumber\\
(M^2_0)_{14} &=  (M^2_0)_{41}= \frac{\mu_I v v_s}{\sqrt{2}v_1}\, N^{-1}_A,  \nonumber\\
(M^2_0)_{22} &=  2 \lambda_2 v_2^2+\frac{\mu_R v_1 v_s}{\sqrt{2}v_2, } \nonumber\\
(M^2_0)_{23} &= (M^2_0)_{32}=2\kappa_2 v_2 v_s-\frac{\mu_R v_1}{\sqrt{2}}+\frac{c_R}{\Lambda^2}\,v^2_1 v_2 v_s, \nonumber\\
(M^2_0)_{24} &= (M^2_0)_{42}= \frac{\mu_I v v_s}{\sqrt{2}v_2}\, N^{-1}_A,  \nonumber\\
(M^2_0)_{33} &=2\lambda_S  v_s^2+\frac{\mu_R  v_1 v_2}{\sqrt{2} v_s},  \nonumber\\
(M^2_0)_{34}  &= (M^2_0)_{43} =  \frac{\mu_I v}{\sqrt{2}}\, N^{-1}_A,  \nonumber\\
(M^2_0)_{44} &= \frac{v^2 v_s}{\sqrt{2}v_1 v_2}\, \Big(\mu_R - \frac{\sqrt{2}c_R v_1 v_2 v_s}{\Lambda^2}\Big)  N^{-2}_A.
  \label{h1matrix}
\end{align}

Following  the procedure in Ref.~\cite{Bian:2017xzg}, we can diagonalize the $3\times 3$ sub-matrix for CP-even scalars in the $4\times 4$ mass matrix in Eq.~(\ref{h1matrix}) as
\begin{equation}
R_h M^2_{3\times 3} R^T_h = {\rm diag} (m^2_{h^0_1},m^2_{h^0_2},m^2_{h^0_3}),
\end{equation}
with the rotation matrix,
\begin{equation}
  R_h =
  \begin{pmatrix}
    c_{\alpha_1} c_{\alpha_2} & s_{\alpha_1} c_{\alpha_2} & s_{\alpha_2}\\
    -(c_{\alpha_1} s_{\alpha_2} s_{\alpha_3} + s_{\alpha_1} c_{\alpha_3})
    & c_{\alpha_1} c_{\alpha_3} - s_{\alpha_1} s_{\alpha_2} s_{\alpha_3}
    & c_{\alpha_2} s_{\alpha_3} \\
    - c_{\alpha_1} s_{\alpha_2} c_{\alpha_3} + s_{\alpha_1} s_{\alpha_3} &
    -(c_{\alpha_1} s_{\alpha_3} + s_{\alpha_1} s_{\alpha_2} c_{\alpha_3})
    & c_{\alpha_2}  c_{\alpha_3}
  \end{pmatrix},
\end{equation}
where $s_{\alpha_i} \equiv \sin\alpha_i$ and $c_{\alpha_i} \equiv \cos\alpha_i$, with $- \pi/2 \leq \alpha_{1,2,3} < \pi/2$.
Then, the mass eigenvalues of CP-even scalars are given by
\begin{align}
    m^2_{h^0_1}&=\frac{1}{2} (a+b -\sqrt{D})\equiv m^2_h, \nonumber \\
    m^2_{h^0_2}&=\frac{1}{2} (a+b+\sqrt{D})\equiv m^2_H,  \nonumber  \\
    m^2_{h^0_3}&=2\lambda_S  v_s^2+\frac{\mu  v_1 v_2}{\sqrt{2} v_s}\equiv
             m^2_s ,  \label{h0s}
\end{align}
where
\begin{equation}
  a\equiv 2 \lambda_1 v_1^2+\frac{\mu_R v_2 v_s}{\sqrt{2} v_1} ,\quad
  b\equiv  2 \lambda_2 v_2^2+\frac{\mu_R v_1 v_s}{\sqrt{2}v_2}, \quad
  D\equiv (a-b)^2+4 d^2,
\end{equation}
with $d\equiv 2 v_1 v_2 (\lambda_3+\lambda_4)+\frac{c_R}{\Lambda^2}\, v_1v_2 v^2_s-\frac{1}{ \sqrt{2}}\mu_R v_s$.
We also denote with $h^0_4\equiv A^0$ that
\begin{equation}
m^2_{h^0_4}\equiv m^2_{A^0}= \frac{v^2 v_s}{\sqrt{2}v_1 v_2}\, \Big(\mu_R - \frac{\sqrt{2}c_R v_1 v_2 v_s}{\Lambda^2}\Big)  N^{-2}_A. \label{h0a}
\end{equation}
The above results have been used for the approximate mass eigenstates in the text.

The charged Goldstone boson $G^+$ and charged
Higgs scalar $H^+$ identified as
\begin{align}
  G^+ &= \cos\beta \, \phi_1^+ + \sin\beta \, \phi_2^+ ,\nonumber\\
  H^+ &= \sin\beta \, \phi_1^+ - \cos\beta \, \phi_2^+ \label{base-chargedH}
\end{align}
with nonzero mass eigenvalue given by
\begin{equation}
  m_{H^+}^2 = m_A^2 - \left ( \frac{\mu\sin\beta \cos\beta}{\sqrt{2}
      v_s} +\lambda_4 \right ) v^2.  \label{H+mass}
\end{equation}

\section{Diagonalization of quark mass matrices and CP phases \label{app:quark_diagonal}}

After two Higgs doublet fields develop VEVs, we obtain the quark mass matrices
from Eqs.~(\ref{qmass1}) and~(\ref{qmass2}) as
\begin{align}
  {(M_u)}_{ij}
  &= \frac{1}{\sqrt{2}}  v\cos\beta
    \begin{pmatrix}
      y^u_{11} & y^u_{12} & 0\\ y^u_{21} & y^u_{22}  & 0 \\ 0 & 0 & y^u_{33}
    \end{pmatrix} + \frac{1}{\sqrt{2}} v\sin\beta \begin{pmatrix}
      0 & 0 & 0\\ 0 & 0  & 0 \\ h^u_{31} & h^u_{32} & 0
    \end{pmatrix},  \nonumber\\
  {(M_d)}_{ij}
  &= \frac{1}{\sqrt{2}}  v\cos\beta
    \begin{pmatrix}
      y^d_{11} & y^d_{12} & 0\\ y^d_{21} & y^d_{22}  & 0 \\ 0 & 0 & y^d_{33}
    \end{pmatrix} + \frac{1}{\sqrt{2}} v\sin\beta \begin{pmatrix}
      0 & 0 & h^d_{13}\\ 0 & 0  & h^d_{23} \\ 0 & 0 & 0
    \end{pmatrix}.
\end{align}
The above quark mass matrices can be diagonalized by
\begin{equation}
  U^\dagger_L M_u U_R
  = M^D_u=
  \diag(m_u, \, m_c, \, m_t),
  \quad
  D^\dagger_L M_d D_R
  = M^D_d=
  \diag(m_d, \, m_s, \, m_b) ,
\end{equation}
thus the CKM matrix is given as $V_\text{CKM}= U^\dagger_L D_L$.
We note that the Yukawa couplings of the second Higgs doublet are
sources of flavor violation, which could be important in meson
decays/mixings and collider searches for flavor-violating top decays
and/or heavy Higgs bosons.

Since $h^u_{31}$ and  $h^u_{32}$ correspond to the rotations of
right-handed up-type quarks, we can take $U_L=1$, so
$V_\text{CKM}=D_L$.
In this case, we have an approximate relation for the down-type quark
mass matrix, $M_d\approx V_\text{CKM} M^D_d $, up to $m_{d,s}/m_b$
corrections. Then the Yukawa couplings between the third and first
two generations are given as follows:
\begin{equation}
  h^d_{13} =\frac{\sqrt{2} m_b}{v\sin\beta}\, V_{ub},\quad
 h^d_{23} =\frac{\sqrt{2} m_b}{v\sin\beta}\, V_{cb}.\label{hd}
\end{equation}
For $V_{ub}\simeq 0.004\ll V_{cb}\simeq 0.04$, we have $h^d_{13}\ll
h^d_{23}$. The down-type Yukawa couplings are determined as
\begin{align}
y^d_{11}&= \frac{\sqrt{2} m_d}{v\cos\beta}\, V_{ud}, \quad
y^d_{12}= \frac{\sqrt{2} m_s}{v\cos\beta}\, V_{us}, \nonumber\\
y^d_{21}&= \frac{\sqrt{2} m_d}{v\cos\beta}\, V_{cd}, \quad
y^d_{22}= \frac{\sqrt{2} m_s}{v\cos\beta}\, V_{cs}, \quad
y^d_{33}= \frac{\sqrt{2} m_b}{v\cos\beta}\, V_{tb}.
\end{align}
We have fixed the down-type Yukawa couplings completely, including the weak CP phase.

Taking $U_L=1$ as above, we find another approximate
relation for the up-type quark mass matrix: $M_u=M^D_u U^\dagger_R$.
Then, the rotation mass matrix for right-handed down-type quarks
becomes $U^\dagger_R={(M^D_u)}^{-1} M_u$, which is given as
\begin{equation}
  U^\dagger_R= \frac{1}{\sqrt{2}}
  \begin{pmatrix}
    \frac{v}{m_u}\cos\beta\,   y^u_{11} &  \frac{v}{m_u}\cos\beta \,   y^u_{12} & 0\\  \frac{v}{m_c}\cos\beta \,   y^u_{21} &  \frac{v}{m_c}\cos\beta \,   y^u_{22}  & 0 \\  \frac{v}{m_t}\sin\beta\,  h^u_{31} &  \frac{v}{m_t}\sin\beta\,  h^u_{32} &  \frac{v}{m_t}\cos\beta\,     y^u_{33}
  \end{pmatrix}.
\end{equation}
From the unitarity condition of $U_R$ we further find the following
constraints on the up-type quark Yukawa couplings:
\begin{align}
  |y^u_{11}|^2 + |y^u_{12}|^2 &= \frac{2m^2_u}{v^2\cos^2\beta}, \label{UR1} \\
  |y^u_{21}|^2+ |y^u_{22}|^2 &=  \frac{2m^2_c}{v^2 \cos^2\beta},  \label{UR1a} 
  \\
  |y^u_{33}|^2+ \tan^2\beta (|h^u_{31}|^2+|h^u_{32}|^2 )&= \frac{2m^2_t}{v^2\cos^2\beta},   \label{UR2}  \\
  y^u_{11}(y^u_{21})^* + y^u_{12} (y^u_{22})^* &= 0,  \label{UR3} \\
   y^u_{21} (h^u_{31})^*+ y^u_{22} (h^u_{32})^* &= 0, \label{UR4}  \\
  y^u_{11} (h^u_{31})^*+ y^u_{12} (h^u_{32})^* &= 0. \label{UR5}
\end{align}

We are now in a position to show the absence of the extra CP phases in the quark Yukawa couplings in our model.
First, performing the simultaneous phase rotations of $q_L$ and $u_R$
as well as $d_R$ to leave the down-type Yukawa couplings untouched, we
can eliminate the CP phases in the diagonal entries, $y^u_{ii}$ with
$i=1$, $2$, $3$. Then, there are four CP phases from $y^u_{12}, y^u_{21}, h^u_{31}, h^u_{32}$, subject to three unitarity conditions in Eqs.~(\ref{UR3})--(\ref{UR5}). Therefore, there remains only one independent CP phase, other than the weak CP phase in the SM\@.
Suppose that off-diagonal entries in the up-type Yukawa matrix are nonzero, so we write $y^u_{21}=e^{i\theta_{21}} |y^u_{21}|$, $y^u_{12}=e^{i\theta_{12}} |y^u_{12}|$, $h^u_{31}=e^{i\theta_{31}} |h^u_{31}|$ and $h^u_{32}=e^{i\theta_{32}} |h^u_{32}|$. Then, Eqs.~(\ref{UR3})--(\ref{UR5}) lead to
\begin{align}
e^{i(\theta_{21}+\theta_{12})}&= -\frac{|y^u_{11} y^u_{21}|}{|y^u_{22} y^u_{12}|}, \label{UR3a} \\
e^{i(\theta_{31}-\theta_{21}-\theta_{32})}&= -\frac{|y^u_{21} h^u_{31}|}{|y^u_{22} h^u_{32}|}, \label{UR4a} \\
e^{i(\theta_{31}+\theta_{12}-\theta_{32})}&= -\frac{|y^u_{11} h^u_{31}|}{|y^u_{12} h^u_{32}|}. \label{UR5a}
\end{align}
But, dividing Eq.~(\ref{UR5a}) by (\ref{UR4a}), we find that $e^{i(\theta_{21}+\theta_{12})}=|y^u_{11} y^u_{22}|/|y^u_{12} y^u_{21}|$. Then, we would get $|y^u_{22}|^2+|y^u_{21}|^2=0$ with Eq.~(\ref{UR3a}), which is inconsistent with Eq.~(\ref{UR1a}).
Therefore, we must choose $y^u_{21}=y^u_{12}=h^u_{31}=h^u_{32}=0$, for which Eqs.~(\ref{UR3})--(\ref{UR5}) are trivially satisfied.
As a result, we find that there is no extra CP phase in the up-type Yukawa couplings either.
Taking $y^u_{21}=y^u_{12}=h^u_{31}=h^u_{32}=0$, Eqs.~(\ref{UR1})--(\ref{UR2}) determine the diagonal down-type Yukawa couplings as
\begin{equation}
 |y^u_{11}|  = \frac{\sqrt{2}m_u}{v\cos\beta}, \quad
 |y^u_{22}| =  \frac{\sqrt{2}m_c}{v\cos\beta},  \quad
  |y^u_{33}|= \frac{\sqrt{2}m_t}{v\cos\beta}.
\end{equation}

\section{Self-interactions and gauge interactions for scalar fields \label{app:higgs_self}}

The couplings for the Higgs self-interactions associated with the charged
Higgs boson are
\begin{align}
  g_{H^+ H^- h_1} =
  &~ ( \lambda_1 c_\alpha s_\beta  + \lambda_2 s_\alpha c_\beta  )
    v s_{2\beta} + 2 \lambda_3 v c_{\beta - \alpha} \nonumber\\
  & - \left( \lambda_3 +
    \lambda_4 + \frac{c_R}{2 \Lambda^2} v_s^2 \right) v s_{\beta + \alpha} s_{2\beta}
  - \frac{\sqrt{2} \mu_I v_s \varepsilon_1}{N_A v} , \nonumber\\
  g_{H^+ H^- h_2} =
  & - ( \lambda_1 s_\alpha s_\beta - \lambda_2 c_\alpha c_\beta)
    v s_{2\beta} + 2 \lambda_3 v s_{\beta - \alpha} \nonumber\\
  & - \left( \lambda_3 + \lambda_4 + \frac{c_R}{2 \Lambda^2} v_s^2
    \right) v c_{\beta + \alpha} s_{2\beta}
  - \frac{\sqrt{2} \mu_I v_s \varepsilon_2}{N_A v} , \nonumber\\
  g_{H^+ H^- h_3} =
  &~ \frac{1}{\sqrt{2}} \mu_R s_{2\beta} + 2 (\kappa_1 s_\beta^2 + \kappa_2
    c_\beta^2) v_s - \frac{c_R}{2\Lambda^2} v^2 v_s s_{2\beta}^2
    - \frac{\sqrt{2} \mu_I v_s \varepsilon_3}{N_A v} ,\nonumber\\
  g_{H^+ H^- h_4}
  =& - g_{H^+ H^- h_1} \varepsilon_1 - g_{H^+ H^- h_2} \varepsilon_2
     - g_{H^+ H^- h_3} \varepsilon_3 - \frac{\sqrt{2} \mu_I v_s}{N_A
     v} + \mathcal{O}(\varepsilon_i^2) .
\end{align}
The quartic couplings can be expressed by the Higgs masses and mixing
angles:
\begin{align}
  \lambda_1
  &= \frac{2 \sum_{i} m_{h_i^0}^2 (R_h)_{i1}^2 - \sqrt{2} \mu_R
    v_s t_\beta}{4 v^2 c_\beta^2}
  \approx \frac{2 (m_{h_1^0}^2 c_\alpha^2 + m_{h_2^0}^2 s_\alpha^2) -
    \sqrt{2} \mu_R v_s t_\beta}{4 v^2 c_\beta^2} ,
    \nonumber\\
  \lambda_2
  &= \frac{2 \sum_{i} m_{h_i^0}^2 (R_h)_{i2}^2 - \sqrt{2} \mu_R
    v_s / t_\beta}{4 v^2 s_\beta^2}
  \approx \frac{2 (m_{h_1^0}^2 s_\alpha^2 + m_{h_2^0}^2 c_\alpha^2) -
    \sqrt{2} \mu_R v_s / t_\beta}{4 v^2 s_\beta^2} ,
    \nonumber\\
  \lambda_3
  &= \frac{\sqrt{2} \mu_R v_s + 2 \sum_i m_{h_i^0}^2 (R_h)_{i1}
    (R_h)_{i2}}{2 v^2 s_{2 \beta}} - \lambda_4 -
    \frac{c_R}{2\Lambda^2} v_s^2
    \nonumber\\
  &\approx \frac{m_{H^+}^2}{v^2} + \frac{s_{2\alpha}}{2s_{2\beta}}
    \frac{m_{h_1^0}^2 - m_{h_2^0}^2}{v^2} - \frac{\mu_R v_s}{\sqrt{2}
    s_{2\beta} v^2} + \frac{c_R}{2\Lambda^2} (v_s^2 + 2 v^2
    s_\beta^2 c_\beta^2) ,
    \nonumber\\
  \lambda_4
  &= - \frac{m_{H^+}^2}{v^2} + \frac{\mu_R v_s}{\sqrt{2} s_\beta
    c_\beta v^2} - \frac{c_R}{\Lambda^2} \frac{v_s^2}{N_A^2} \quad
    \nonumber\\
  \kappa_1 + \frac{c_R}{2\Lambda^2} v^2 s_\beta^2
  &= \frac{\sqrt{2} \mu_R v s_\beta + 2 \sum_i m_{h_i^0}^2 (R_h)_{i1}
    (R_h)_{i3}}{4 v v_s c_\beta}
    \approx \frac{\mu_R s_\beta}{2 \sqrt{2} v_s c_\beta} ,
    \nonumber\\
  \kappa_2 + \frac{c_R}{2\Lambda^2} v^2 c_\beta^2
  &= \frac{\sqrt{2} \mu_R v c_\beta + 2 \sum_i m_{h_i^0}^2 (R_h)_{i2}
    (R_h)_{i3}}{4 v v_s s_\beta}
    \approx \frac{\mu_R c_\beta}{2 \sqrt{2} v_s s_\beta} .
\end{align}
Here we have taken the limit where the mixing with the singlet field is
negligible.
Using the relations in the above, the Higgs couplings are now given as
\begin{align}
  g_{H^+ H^- h_1}
  \approx
  &~ c_{\beta - \alpha} \frac{2 m_{H^+}^2}{v} +
    \left( \frac{s_\alpha c_\beta^2}{s_\beta} + \frac{c_\alpha
    s_\beta^2}{c_\beta} \right) \frac{m_{h_1^0}^2}{v} \nonumber\\
  & - \frac{\mu_R v_s s_{\beta + \alpha} }{\sqrt{2} v s_\beta^2
    c_\beta^2}
    - \frac{\sqrt{2} \mu_I v_s \varepsilon_1}{N_A v}
    + \frac{c_R}{\Lambda^2} v (v_s^2 + 2 v^2 s_\beta^2 c_\beta^2)
    c_{\beta - \alpha} ,
    \\
  g_{H^+ H^- h_2}
  \approx
  &~ s_{\beta - \alpha} \frac{2 m_{H^+}^2}{v} +
    \left( \frac{c_\alpha c_\beta^2}{s_\beta} - \frac{s_\alpha
    s_\beta^2}{c_\beta} \right) \frac{m_{h_2^0}^2}{v} \nonumber\\
  & - \frac{\mu_R v_s c_{\beta + \alpha} }{\sqrt{2} v s_\beta^2
    c_\beta^2}
    - \frac{\sqrt{2} \mu_I v_s \varepsilon_2}{N_A v}
    + \frac{c_R}{\Lambda^2} v (v_s^2 + 2 v^2 s_\beta^2 c_\beta^2)
    s_{\beta - \alpha} ,
    \\
  g_{H^+ H^- h_3} \approx
  &~ \frac{\mu_R}{\sqrt{2} s_\beta c_\beta} - \frac{\sqrt{2} \mu_I v_s
    \varepsilon_3}{N_A v}
    - \frac{c_R}{\Lambda^2} v^2 v_s  .
\end{align}

The Higgs interactions to the $W$ bosons arise through the kinetic
terms:
\begin{align}
  \mathcal{L}_K
  =&~ \abs{D_\mu H_1}^2 + \abs{D_\mu H_2}^2 \nonumber\\
  \supset
   &~ \frac{g^2v}{2} (c_\beta \rho_1 + s_\beta \rho_2) W_\mu^+
    W^{- \mu}
    + \left[ \frac{i g}{2} W_\mu^+ \left(
    \phi_1^- \partial^\mu \rho_1 - \rho_1 \partial^\mu \phi_1^-
    + \phi_2^- \partial^\mu \rho_2 - \rho_2 \partial^\mu \phi_2^-
     \right) \right .
     \nonumber\\
   & \left .
     - \frac{g}{2} W_\mu^+ \left(
    \phi_1^- \partial^\mu \eta_1 - \eta_1 \partial^\mu \phi_1^-
    + \phi_2^- \partial^\mu \eta_2 - \eta_2 \partial^\mu \phi_2^-
     \right) + \mathrm{c.c.}
     \right] \nonumber\\
  \approx
  &~\frac{2 m_W^2}{v} \bigg[ c_{\beta - \alpha} h_1  + s_{\beta
    - \alpha} h_2 - (c_{\beta  - \alpha} \varepsilon_1 + s_{\beta -
    \alpha} \varepsilon_2) h_4
    \bigg] W_\mu^+ W^{- \mu} \nonumber\\
   & + \left[ \frac{i g (s_{\beta - \alpha} + i N_A \varepsilon_1)}{2} W_\mu^+
     \left( H^- \partial^\mu h_1 - h_1 \partial^\mu H^- \right)
     \right . \nonumber\\
  &\quad
     - \frac{i g (c_{\beta - \alpha} - i N_A \varepsilon_2)}{2} W_\mu^+
     \left( H^- \partial^\mu h_2 - h_2 \partial^\mu H^- \right)
    \nonumber\\
   &\quad
     - \frac{g N_A \varepsilon_3}{2} W_\mu^+
     \left( H^- \partial^\mu h_3 - h_3 \partial^\mu H^- \right)
     \nonumber\\
   &\quad \left .
     - \frac{i g (s_{\beta - \alpha} \varepsilon_1 - c_{\beta -
     \alpha} \varepsilon_2 - i N_A)}{2} W_\mu^+
     \left( H^- \partial^\mu h_4 - h_4 \partial^\mu H^- \right)
     + \mathrm{c.c.} \right] .
\end{align}
Note that the charged Higgs boson interacts with the neutral Higgs and
$W$ bosons through derivative couplings. The couplings are
\begin{align}
  i g_{h_1 W^\pm H^\mp}^\mu
  &= - \frac{g}{2} \left( N_A \varepsilon_1 \mp i s_{\beta - \alpha}
    \right) (p_{h_1} - p_{H^\pm})^\mu , \nonumber\\
  i g_{h_2 W^\pm H^\mp}^\mu
  &= -\frac{g}{2} \left( N_A \varepsilon_2 \pm i c_{\beta - \alpha}
    \right)
    (p_{h_2} - p_{H^\pm})^\mu, \nonumber\\
  i g_{h_3 W^\pm H^\mp}^\mu
  &= - \frac{g N_A \varepsilon_3}{2} (p_{h_3} - p_{H^\pm})^\mu ,
    \nonumber\\
  i g_{h_4 W^\pm H^\mp}^\mu
  &= - \frac{g}{2} \left[ N_A \pm i \Big( s_{\beta - \alpha} \varepsilon_1
    - c_{\beta - \alpha} \varepsilon_2 \right) \Big] (p_{h_4}
    - p_{H^\pm})^\mu .
\end{align}
Here all the momenta $p_{h_i}$ and $p_{H^\pm}$ are all incoming
to the vertices.

Similarly, the Higgs interactions to the $Z$ bosons are given from the following terms:
\begin{align}
  \mathcal{L}_K
   \supset&~
   \frac{m^2_Z}{2v^2}\Big[ (v_1+\rho_1)^2 + (v_2+ \rho_2)^2 \Big] Z_\mu
    Z^{ \mu} + \frac{m_Z}{v} Z_\mu \Big( \rho_1 \partial^\mu \eta_1 -
    \eta_1 \partial^\mu \rho_1 + \rho_2 \partial^\mu \eta_2 - \eta_2
    \partial^\mu \rho_2 \Big)
    \nonumber\\
  \approx&~  \frac{m^2_Z}{v} \bigg[ c_{\beta-\alpha} h_1 + s_{\beta-\alpha} h_2-(\varepsilon_1 c_{\beta-\alpha}+\varepsilon_2 s_{\beta-\alpha})h_4\bigg] Z_\mu Z^\mu \nonumber \\
  &+ \frac{m^2_Z}{2v^2}\Big[h^2_1+h^2_2-2\varepsilon_1 h_1h_4- 2\varepsilon_2 h_2h_4 \Big]Z_\mu Z^\mu \nonumber \\
   & + \frac{N_A m_Z}{v} Z_\mu \Big[ s_{\beta - \alpha} \left(
    h_1 \partial^\mu h_4 - h_4 \partial^\mu h_1 \right)
    - c_{\beta - \alpha} \left( h_2 \partial^\mu h_4 - h_4
    \partial^\mu h_2 \right) \nonumber\\
  &\qquad\qquad\qquad
    + \varepsilon_3 s_{\beta - \alpha} \left( h_1 \partial^\mu h_3
    - h_3 \partial^\mu h_1 \right)
    - \varepsilon_3 c_{\beta - \alpha} \left( h_2 \partial^\mu h_3
    - h_3 \partial^\mu h_2 \right) \nonumber\\
  &\qquad\qquad\qquad
    - ( c_{\beta - \alpha} \varepsilon_1 - s_{\beta - \alpha}
    \varepsilon_2) \left( h_1 \partial^\mu h_2
    - h_2 \partial^\mu h_1 \right) \Big] + \mathcal{O}(\varepsilon_i^2).
\end{align}


\begin{thebibliography}{999}


\bibitem{2HDM}
G.~C.~Branco, P.~M.~Ferreira, L.~Lavoura, M.~N.~Rebelo, M.~Sher and J.~P.~Silva,
Phys. Rept. \textbf{516} (2012), 1-102
[arXiv:1106.0034 [hep-ph]].




\bibitem{RK}
  R.~Aaij {\it et al.} [LHCb Collaboration],
  Phys.\ Rev.\ Lett.\  {\bf 113} (2014) 151601
  [arXiv:1406.6482 [hep-ex]].

\bibitem{RK-new}
R.~Aaij \textit{et al.} [LHCb],
Phys. Rev. Lett. \textbf{122}, no.19, 191801 (2019)
[arXiv:1903.09252 [hep-ex]].


\bibitem{RKs}
S. Bifani (2017), Seminar at CERN, \url{https://indico.cern.ch/event/580620/};
  S.~Bifani [LHCb Collaboration],
  arXiv:1705.02693 [hep-ex];
  R.~Aaij {\it et al.} [LHCb Collaboration],
  JHEP {\bf 1708} (2017) 055
  [arXiv:1705.05802 [hep-ex]].



\bibitem{P5}
  R.~Aaij {\it et al.} [LHCb Collaboration],
  Phys.\ Rev.\ Lett.\  {\bf 111} (2013) 191801
  [arXiv:1308.1707 [hep-ex]];
  R.~Aaij {\it et al.} [LHCb Collaboration],
  JHEP {\bf 1602} (2016) 104
  [arXiv:1512.04442 [hep-ex]].


\bibitem{RKs-new}
A.~Abdesselam \textit{et al.} [Belle],
arXiv:1904.02440 [hep-ex].


\bibitem{altman}
W.~Altmannshofer, P.~S.~B.~Dev, A.~Soni and Y.~Sui,
Phys. Rev. D \textbf{102}, no.1, 015031 (2020)
[arXiv:2002.12910 [hep-ph]].



\bibitem{RK-newfit}
J.~Aebischer, W.~Altmannshofer, D.~Guadagnoli, M.~Reboud, P.~Stangl and D.~M.~Straub,
Eur. Phys. J. C \textbf{80} (2020) no.3, 252
[arXiv:1903.10434 [hep-ph]].


\bibitem{update}
M.~Alguer{\'o}, B.~Capdevila, A.~Crivellin, S.~Descotes-Genon, P.~Masjuan, J.~Matias, M.~Novoa Brunet and J.~Virto,
Eur. Phys. J. C \textbf{79} (2019) no.8, 714
[arXiv:1903.09578 [hep-ph]].



\bibitem{Bian:2017rpg}
  L.~Bian, S.~M.~Choi, Y.~J.~Kang and H.~M.~Lee,
  Phys.\ Rev.\ D {\bf 96} (2017) no.7,  075038
  [arXiv:1707.04811 [hep-ph]].

\bibitem{Bian:2017xzg}
  L.~Bian, H.~M.~Lee and C.~B.~Park,
  Eur.\ Phys.\ J.\ C {\bf 78} (2018) no.4,  306
  [arXiv:1711.08930 [hep-ph]].


\bibitem{b3l3}
K.~S.~Babu, A.~Friedland, P.~A.~N.~Machado and I.~Mocioiu,
JHEP \textbf{12} (2017), 096
[arXiv:1705.01822 [hep-ph]];
R.~Alonso, P.~Cox, C.~Han and T.~T.~Yanagida,
Phys. Lett. B \textbf{774} (2017), 643-648
[arXiv:1705.03858 [hep-ph]];
P.~Ko, T.~Nomura and C.~Yu,
JHEP \textbf{04} (2019), 102
[arXiv:1902.06107 [hep-ph]].


\bibitem{lmultau}
W.~Altmannshofer, S.~Gori, M.~Pospelov and I.~Yavin,
Phys. Rev. D \textbf{89} (2014), 095033
[arXiv:1403.1269 [hep-ph]];
A.~Crivellin, G.~D'Ambrosio and J.~Heeck,
Phys. Rev. D \textbf{91} (2015) no.7, 075006
[arXiv:1503.03477 [hep-ph]].

\bibitem{CPansatz}
S.~Kanemura, M.~Kubota and K.~Yagyu,
JHEP \textbf{08}, 026 (2020)
[arXiv:2004.03943 [hep-ph]].


\bibitem{Aaij:2021vac}
R.~Aaij \textit{et al.} [LHCb],
arXiv:2103.11769 [hep-ex].

\bibitem{RK-belle-new}
A.~Abdesselam \textit{et al.} [BELLE],
JHEP \textbf{03}, 105 (2021)
[arXiv:1908.01848 [hep-ex]].




\bibitem{babar}
  J.~P.~Lees {\it et al.} [BaBar Collaboration],
  Phys.\ Rev.\ Lett.\  {\bf 109} (2012) 101802
  [arXiv:1205.5442 [hep-ex]];
  J.~P.~Lees {\it et al.} [BaBar Collaboration],
  Phys.\ Rev.\ D {\bf 88} (2013) no.7,  072012
  [arXiv:1303.0571 [hep-ex]].

 \bibitem{belle}
  M.~Huschle {\it et al.} [Belle Collaboration],
  Phys.\ Rev.\ D {\bf 92} (2015) no.7,  072014
  [arXiv:1507.03233 [hep-ex]];
  A.~Abdesselam {\it et al.} [Belle Collaboration],
  arXiv:1603.06711 [hep-ex].

\bibitem{belle-new1}
A.~Abdesselam \textit{et al.} [Belle],
[arXiv:1904.08794 [hep-ex]].


\bibitem{lhcb}
  R.~Aaij {\it et al.} [LHCb Collaboration],
  Phys.\ Rev.\ Lett.\  {\bf 115} (2015) no.11,  111803  
  [arXiv:1506.08614 [hep-ex]]; Erratum: [Phys.\ Rev.\ Lett.\  {\bf 115} (2015) no.15,  159901]


\bibitem{hflav}
Y.~S.~Amhis \textit{et al.} [HFLAV],
arXiv:1909.12524 [hep-ex].



\bibitem{belle-new}
G.~Caria \textit{et al.} [Belle],
Phys. Rev. Lett. \textbf{124} (2020) no.16, 161803
[arXiv:1910.05864 [hep-ex]].


\bibitem{LQ}
S.~M.~Choi, Y.~J.~Kang, H.~M.~Lee and T.~G.~Ro,
JHEP \textbf{10} (2018), 104
[arXiv:1807.06547 [hep-ph]].


\bibitem{allanach}
B.~C.~Allanach, J.~M.~Butterworth and T.~Corbett,
JHEP \textbf{08} (2019), 106
[arXiv:1904.10954 [hep-ph]].





\bibitem{acme}
  V.~Andreev {\it et al.} [ACME Collaboration],
  Nature {\bf 562} (2018) no.7727,  355.


\bibitem{Barr:1990vd}
  S.~M.~Barr and A.~Zee,
  Phys.\ Rev.\ Lett.\  {\bf 65}, 21 (1990)
  Erratum: [Phys.\ Rev.\ Lett.\  {\bf 65}, 2920 (1990)].


\bibitem{Chang:1990sf}
  D.~Chang, W.~Y.~Keung and T.~C.~Yuan,
  Phys.\ Rev.\ D {\bf 43}, 14 (1991).



\bibitem{Leigh:1990kf}
  R.~G.~Leigh, S.~Paban and R.~M.~Xu,
  Nucl.\ Phys.\ B {\bf 352}, 45 (1991).


\bibitem{Abe:2013qla}
  T.~Abe, J.~Hisano, T.~Kitahara and K.~Tobioka,
  JHEP {\bf 1401}, 106 (2014)
  Erratum: [JHEP {\bf 1604}, 161 (2016)]
  [arXiv:1311.4704 [hep-ph]].

\bibitem{Shu:2013uua}
J.~Shu and Y.~Zhang,
Phys. Rev. Lett. \textbf{111} (2013) no.9, 091801
[arXiv:1304.0773 [hep-ph]].

\bibitem{Bian:2014zka}
L.~Bian, T.~Liu and J.~Shu,
Phys. Rev. Lett. \textbf{115} (2015), 021801
[arXiv:1411.6695 [hep-ph]].

\bibitem{Bian:2016zba}
L.~Bian and N.~Chen,
Phys. Rev. D \textbf{95} (2017) no.11, 115029
[arXiv:1608.07975 [hep-ph]].

\bibitem{Bian:2017jpt}
L.~Bian, N.~Chen and Y.~Zhang,
Phys. Rev. D \textbf{96} (2017) no.9, 095008
[arXiv:1706.09425 [hep-ph]].



\bibitem{amu}
  G.~W.~Bennett {\it et al.} [Muon g-2 Collaboration],
  Phys.\ Rev.\ D {\bf 73} (2006) 072003
  [hep-ex/0602035].




\bibitem{pdg}
  C.~Patrignani {\it et al.} [Particle Data Group],
  Chin.\ Phys.\ C {\bf 40} (2016) no.10,  100001.


\bibitem{g2update}
  T.~Aoyama, \textit{et al.}
Phys. Rept. \textbf{887}, 1-166 (2020)
[arXiv:2006.04822 [hep-ph]].




\bibitem{Haisch:2018djm}
  U.~Haisch and G.~Polesello,
  JHEP \textbf{09}, 151 (2018)
  [arXiv:1807.07734 [hep-ph]].

\bibitem{Kling:2018xud}
  F.~Kling, H.~Li, A.~Pyarelal, H.~Song and S.~Su,
  JHEP \textbf{06}, 031 (2019)
  [arXiv:1812.01633 [hep-ph]].

\bibitem{Haber:1992py}
  H.~E.~Haber and A.~Pomarol,
  Phys. Lett. B \textbf{302}, 435-441 (1993)
  [arXiv:hep-ph/9207267 [hep-ph]].

\bibitem{Pomarol:1993mu}
  A.~Pomarol and R.~Vega,
  Nucl. Phys. B \textbf{413}, 3-15 (1994)
  [arXiv:hep-ph/9305272 [hep-ph]].

\bibitem{Gerard:2007kn}
  J.~M.~Gerard and M.~Herquet,
  Phys. Rev. Lett. \textbf{98}, 251802 (2007)
  [arXiv:hep-ph/0703051 [hep-ph]].

\bibitem{Grzadkowski:2010dj}
  B.~Grzadkowski, M.~Maniatis and J.~Wudka,
  JHEP \textbf{11}, 030 (2011)
  [arXiv:1011.5228 [hep-ph]].

\bibitem{Haber:2010bw}
  H.~E.~Haber and D.~O'Neil,
  Phys. Rev. D \textbf{83}, 055017 (2011)
  [arXiv:1011.6188 [hep-ph]].

\bibitem{Kling:2020hmi}
  F.~Kling, S.~Su and W.~Su,
  JHEP \textbf{06}, 163 (2020)
  [arXiv:2004.04172 [hep-ph]].

\bibitem{Aaboud:2017cxo}
  M.~Aaboud \textit{et al.} [ATLAS],
  JHEP \textbf{03}, 174 (2018)
  [arXiv:1712.06518 [hep-ex]].

\bibitem{Aaboud:2018eoy}
  M.~Aaboud \textit{et al.} [ATLAS],
  Phys. Lett. B \textbf{783}, 392-414 (2018)
  [arXiv:1804.01126 [hep-ex]].

\bibitem{Sirunyan:2019xls}
  A.~M.~Sirunyan \textit{et al.} [CMS],
  Eur. Phys. J. C \textbf{79}, no.7, 564 (2019)
  [arXiv:1903.00941 [hep-ex]].

\bibitem{Sirunyan:2019xjg}
  A.~M.~Sirunyan \textit{et al.} [CMS],
  JHEP \textbf{03}, 065 (2020)
  [arXiv:1910.11634 [hep-ex]].


\bibitem{UMSSM}
D.~A.~Demir and L.~Everett,
Phys. Rev. D \textbf{69} (2004), 015008
[arXiv:hep-ph/0306240 [hep-ph]].

\bibitem{Bian:2017wfv}
L.~Bian, H.~K.~Guo and J.~Shu,
Chin. Phys. C \textbf{42} (2018) no.9, 093106
[arXiv:1704.02488 [hep-ph]].


\bibitem{Zhou:2020xqi}
R.~Zhou and L.~Bian,
arXiv:2001.01237 [hep-ph].


\bibitem{Dorsch:2016nrg}
G.~C.~Dorsch, S.~J.~Huber, T.~Konstandin and J.~M.~No,
JCAP \textbf{05} (2017), 052
[arXiv:1611.05874 [hep-ph]].



\end{thebibliography}
\end{document}